\documentclass[aps,amsmath,nofootinbib,11pt,eqsecnum]{revtex4}

\usepackage{verbatim}
\usepackage{bm}
\usepackage{epsfig}
\usepackage{psfrag}

\def\Nc{N_{\rm c}}
\def\Nf{N_{\rm f}}
\def\k{{\bm k}}
\def\al{\alpha_{\rm\scriptscriptstyle EM}}

\def\alphaSYM{\alpha_{\rm\scriptscriptstyle SYM}}
\def\J{J^{\rm\scriptscriptstyle EM}}
\def\Ep{{\bm E}_\perp}
\def\wn{w} 
\def\qn{q} 

\def\chimm{\chi^\mu_{\ \mu}}
\def\A{{\cal A}}
\def\B{{\cal B}}

\def\Nfour{\mathcal N\,{=}\,4}

\def\putbox#1#2{\epsfxsize=#1\textwidth\epsfbox{#2}}
\def\centerbox#1#2{\centerline{\epsfxsize=#1\textwidth\epsfbox{#2}}}
\def\sumint{{\textstyle \sum} \hspace{-1.2em}\int}
\def\sgn{{\rm sgn}}
\def\Tr{\,{\rm Tr}\:}

\def\st{\begin{equation}}
\def\stp{\end{equation}}

\def\lra{\leftrightarrow}
\def\Re{\,{\rm Re}\:}
\def\Im{\,{\rm Im}\:}
\def\Eq#1{Eq.~(\ref{#1})}
\def\twotwo{{2\lra2}}

\def\p{{\bm p}}
\def\q{{\bm q}}
\def\k{{\bm k}}

\def\f{{\bm f}}

\def\nf{N_{\rm f}\,}
\def\nc{N_{\rm c}\,}
\def\ncs{N_{\rm c}^2}
\def\mD{m_{\rm D}}

\def\nbose{n_{\rm b}}
\def\nfermi{n_{\rm f}}

\def\half{{\textstyle{\frac 12}}}
\def\coeff#1#2{{\textstyle{\frac {#1}{#2}}}}

\def\alphas{\alpha_{\rm s}}
\def\alphaEM{\alpha_{\rm EM}}

\def\nott#1{\setbox0=\hbox{$#1$}                
   \dimen0=\wd0                                 
   \setbox1=\hbox{/} \dimen1=\wd1               
   \ifdim\dimen0>\dimen1                        
      \rlap{\hbox to \dimen0{\hfil/\hfil}}      
      #1                                        
   \else                                        
      \rlap{\hbox to \dimen1{\hfil$#1$\hfil}}   
      /                                         
   \fi}                                         %

\def\gsim{\mbox{~{\raisebox{0.4ex}{$>$}}\hspace{-1.1em}
	{\raisebox{-0.6ex}{$\sim$}}~}}
\def\lsim{\mbox{~{\raisebox{0.4ex}{$<$}}\hspace{-1.1em}
	{\raisebox{-0.6ex}{$\sim$}}~}}

\def\ofo{ { {}_{2\!}F_1 }}

\advance\parskip 2pt

\begin{document}

\vspace*{-1.5cm}

\hfill {NSF-KITP-06-49}
\medskip

\title{Photon and dilepton production in supersymmetric Yang-Mills plasma}

\author{Simon C.~Huot}
\email{sicah@physics.mcgill.ca}
\affiliation{Department of Physics, McGill University,
 Montreal, QC H3A 2T8, Canada }
\author{Pavel Kovtun}
\email{kovtun@kitp.ucsb.edu}
\affiliation{KITP, University of California, Santa Barbara, CA
 93106-4030, USA}%
\author{Guy D.~Moore}
\email{guymoore@physics.mcgill.ca}
\affiliation{Department of Physics, McGill University,
 Montreal, QC H3A 2T8, Canada}
\author{Andrei Starinets}
\email{starina@perimeterinstitute.ca}
\affiliation{Perimeter Institute for Theoretical Physics,
  Waterloo, ON N2L 2Y5, Canada}
\author{Laurence G.~Yaffe}
\email{yaffe@phys.washington.edu}
\affiliation{Department of Physics, University of Washington,
 Seattle, WA, 98195-1560, USA}

\date{July 2006\\[20pt]}

\begin{abstract}
\noindent
By weakly gauging one of the $U(1)$ subgroups of the
$R$-symmetry group, $\Nfour$ super-Yang-Mills theory
can be coupled to electromagnetism, thus allowing a computation of photon
production and related phenomena in a QCD-like non-Abelian plasma
at both weak and strong coupling.
We compute photon and dilepton emission rates 
from finite temperature $\Nfour$ supersymmetric Yang-Mills plasma
both perturbatively at weak coupling to leading order,
and non-perturbatively at strong coupling
using the AdS/CFT duality conjecture.
Comparison of the photo-emission spectra for $\Nfour$ plasma
at weak coupling, $\Nfour$ plasma at strong coupling,
and QCD at weak coupling reveals several systematic trends which we discuss.
We also evaluate the electric conductivity of $\Nfour$ plasma
in the strong coupling limit, and to leading-log order at weak coupling.
Current-current spectral functions in the strongly coupled theory
exhibit hydrodynamic peaks at small frequency,
but otherwise show no structure which
could be interpreted as well-defined thermal resonances
in the high-temperature phase.
\end{abstract}

\maketitle

\section{Introduction}
Any thermal medium composed of electrically charged particles
emits photons.
The energy spectrum of the produced photons 
depends on the details of the system:
the spectrum is Planckian when the photons are
in thermal equilibrium, and deviates from it when they are not.
The quark-gluon plasma (QGP) produced in heavy ion collisions is expected to
be optically thin, because of its limited extent and the small value
of the electromagnetic coupling $\al$.
Therefore, a photon, once emitted, should stream through the QCD plasma
virtually without subsequent interaction \cite{photons-rhic}.
In such a situation, the photon spectrum will have little to do with
the black-body distribution, but may instead give valuable information
about the properties of the medium.  For instance, while the
experimental results \cite{RHIC_photon1,RHIC_photon2}
for photon production at RHIC
are currently consistent with pion decay plus prompt photons produced by
the initial scattering of the partons from the two nuclei
\cite{Vogelsang}, there is room for 
photons produced in the hot plasma.

While prompt photons really are perturbative, and pion decay photons can
be calibrated from other hadronic signals, the most interesting signal,
photon production from the medium, suffers from the usual problem that
we only have weak coupling calculations
\cite{AMY3}
for QGP photon production, despite the fact that
the medium is probably strongly coupled.
Therefore, any guidance on the behavior of photon production as a
function of coupling would be useful, even if it comes from an analogue
theory which is not quite real QCD.  With this in mind, we will calculate
photon production in $\Nfour$ supersymmetric Yang-Mills theory,
where strong coupling techniques exist.

Consider a field theory in thermal equilibrium, and
let the photon interaction with matter be of the
form $e\J_\mu A^\mu$, where the electromagnetic coupling $e$
is so small that the photons are not rescattered and thermalized.
If $\Gamma_\gamma$ denotes the number of photons emitted per unit time
per unit volume, then to leading order in $e$ the rate
is given by \cite{Le-Bellac}%
\begin{equation}
 d\Gamma_\gamma = \frac{d^3k}{(2\pi)^3} \frac{e^2}{2|\k|} \;
            \eta^{\mu\nu} C^<_{\mu\nu}(K)\Big|_{k^0=|\k|}\ ,
\label{eq:photon-rate-general}
\end{equation}
where
\begin{equation}
  C^<_{\mu\nu}(K) = \int\!d^4X\> e^{-iK\cdot X} \>
  \langle \J_\mu(0)\J_\nu(X)\rangle 
\label{eq:Wightman}
\end{equation}
is the Wightman function of electromagnetic currents,
and the expectation value is taken in the thermal equilibrium state.
Here $\eta_{\mu\nu}={\rm diag}({-}{+}{+}{+})$
is the Minkowski metric, and $K \equiv (k^0,\k)$ is a null 4-vector%
\footnote{%
    We follow the common thermal field theory convention that 4-vectors are
    capitalized while their components are lower case.
    }
whose time component is fixed by the on-shell condition $k^0=|\k|$.
The Wightman correlator (\ref{eq:Wightman}), in thermal equilibrium,
is related to the spectral density,
\st
  C^<_{\mu\nu}(K) = 
  \nbose (k^0)\,\chi_{\mu\nu}(K) \,,
\label{KMS}
\stp
where $\nbose (k^0)=1/(e^{\beta k^0}{-}1)$
is the usual Bose-Einstein distribution function,
and $\chi_{\mu\nu}(K)$
is the spectral density, proportional to the imaginary part
of the retarded current-current correlation function,
\begin{equation}
\chi_{\mu\nu}(K)=-2\,{\rm Im}\, C^{\rm ret}_{\mu\nu}(K) \,.
\end{equation}

If one also adds to the theory massive fermions which carry
only electric charge (``leptons''),
then the thermal system will also emit these leptons, produced
by virtual photon decay.
The same electromagnetic current-current correlation function,
evaluated for timelike momenta,
gives the rate of lepton pair production for each such lepton species
\cite{Le-Bellac}:
\begin{equation}
  d\Gamma_{\ell\bar\ell} = \frac{d^4 K}{(2\pi)^4} \,
  \frac{e^2 \, e_\ell^2}{6\pi |K^2|^{5/2}} \>
  \Theta(k^0)
  \Theta(-K^2{-}4m^2) \,
  [-K^2{-}{4m^2}]^{1/2} \,
  (-K^2{+}{2m^2}) \;
  \eta^{\mu\nu} C^<_{\mu\nu}(K)\,.
\label{eq:dilepton-rate-general}
\end{equation}
Here $e_\ell$ is the electric charge of the lepton,
$m$ is lepton mass, and the correlator
$C^<_{\mu\nu}(K)$ is evaluated at the timelike
momentum of the emitted particle pair.
[$\Theta(x)$ denotes a unit step function.]
Expressions (\ref{eq:photon-rate-general}) and
(\ref{eq:dilepton-rate-general}) for the production rates
are true to leading order in the electromagnetic couplings
$e$ and $e_\ell$,
but are valid non-perturbatively in all other interactions.

The electrical conductivity $\sigma$ of the medium is also
determined by the current-current correlator,
specifically its zero-frequency limit at vanishing three-momentum,%
\footnote
    {
    This Kubo formula for the conductivity is more commonly written in
    terms of the purely spatial part of the correlator,
    $
    \sigma = \lim_{k^0\to0} \frac{e^2}{6T}\, C_{ii}^<(k^0,\k{=}0)
    $.
    The form (\ref{eq:Kubo-sigma}) is equivalent since transversality
    of the current-current correlator implies that
    $ C^<_{00}(k^0,\k{=}0) = 0 $ for any non-zero frequency.
    }
\begin{equation}
  \sigma
    = \lim_{k^0\to0} \frac{e^2}{6T}\>\eta^{\mu\nu} C_{\mu\nu}^<(k^0,\k{=}0)\,.
\label{eq:Kubo-sigma}
\end{equation}
An equally valid alternative expression relates
the conductivity to the the small frequency limit of the
correlator for lightlike momenta,
\begin{equation}
    \sigma = \lim_{k^0 \rightarrow 0} \frac{e^2}{4T} \>
    \eta^{\mu\nu}C^<_{\mu\nu}(K) \Big|_{|\k|=k^0} \,.
\label{eq:Kubo2}
\end{equation}
This form, which will be useful in our discussion of the
photon production rate,
follows from the Ward identity $K^\mu C^<_{\mu\nu}(K) = 0$
combined with the diffusive nature of the hydrodynamic pole
in the correlator.

Both photon and dilepton production rates have previously been calculated in
QCD in perturbation theory to leading order in $\alphas$
[{\em i.e.}, up to relative corrections suppressed by powers of $\alphas$]
\cite{AMY3,rates-pert2}.
There are also non-perturbative lattice estimates of the dilepton emission
rate at zero three-momentum \cite{rates-lattice}, and
of the electric conductivity \cite{Gupta},
based on attempts to fit the Euclidean correlator
using parameterized forms of the spectral density.
Reports of lattice studies of current-current spectral functions
at non-zero momentum have appeared recently \cite{hep-lat/0607012}.
(However, the results of these efforts to extract
real-time physics from Euclidean lattice simulations 
are quite sensitive to the assumptions made about the form of the
spectral density.
Assessing the reliability of these results is not easy;
see, for example, Ref.~\cite{rates-review}.)

In this paper, we calculate photon and dilepton production rates in
$SU(\Nc)$, $\Nfour$ supersymmetric Yang-Mills (SYM) theory
at finite temperature and zero chemical potential,
at both weak and strong coupling.
This theory, at non-zero temperature,
mimics many features of high-temperature QCD.
It is a non-Abelian plasma which happens to have adjoint representation
fermions and scalars instead of fundamental representation quarks.
Despite this difference in matter field content, thermal SYM theory
exhibits deconfinement, Debye screening,
area-law behavior of spatial Wilson loops,
and a finite static correlation length, just like hot QCD.
But unlike QCD, real-time thermal properties of $\Nfour$ SYM theory
can be studied analytically at strong coupling.%
\footnote{
  $\Nfour$ SYM theory is a conformal field theory, whose coupling
  is a fixed, scale-independent parameter.
  The accessibility of the strong coupling regime in SYM theory is due to
  gauge-string duality \cite{AGMOO}, commonly referred to as
  AdS/CFT correspondence.
  Though not proven rigorously, this duality has survived an impressive
  number of consistency tests and
  we assume its validity.
}
Consequently, in SYM theory one may calculate, reliably,
interesting physical observables at both weak and strong coupling.
In particular,
the calculation of thermal spectral functions in strongly coupled
SYM theory is vastly simpler than the corresponding problem
in strongly coupled QCD.
Full spectral functions of the energy-momentum tensor, at strong coupling,
were calculated recently in
Refs.~\cite{spectral,derek},
explicitly showing that strongly coupled SYM theory
behaves much more like a liquid, rather than a weakly
interacting gas of quasi-particles.
At the same time, weak-coupling calculations of the photon emission rates
in SYM theory are qualitatively similar to those in QCD,
and detailed comparison of the results can shed light on the degree to which
thermal QCD can be quantitatively modeled by $\Nfour$ SYM theory.

Our paper is organized as follows.
In section \ref{sec:SYM-EM} we discuss
how to couple the degrees of freedom of $\Nfour$
SYM theory to electromagnetism.
In section \ref{sec:strong-coupling} we calculate the trace of the
spectral function 
$\chimm(K)\equiv\eta^{\mu\nu}\chi_{\mu\nu}(K)$
in strongly coupled SYM theory for arbitrary momenta.
The behavior of the spectral functions at small frequencies
is in complete agreement with the prediction of linear response for
hydrodynamic fluctuations of the conserved charge density.
We find a finite result,
$\sigma=e^2 \Nc^2 T/16\pi$,
for strong coupling limit of the electric conductivity of $\Nfour$ SYM theory.
In section \ref{sec:weak-coupling} we compute the current-current
spectral function in weakly coupled SYM theory, for both timelike and
lightlike momenta,
for frequencies large compared to $\lambda^2 \, T$,
which is the scale where hydrodynamic effects become important.
(Here $\lambda \equiv g^2 \nc$ is the 't~Hooft coupling.)
This weak-coupling analysis generalizes the corresponding calculation
for QCD performed by Arnold, Moore and Yaffe \cite{AMY3}.
In the final section \ref{sec:discussion} we compare the photon emission
spectra for SYM and QCD at weak coupling, and SYM at strong coupling,
and discuss the relevant lessons which can be drawn.
We find that the weak-coupling behavior of SYM is quite similar to that of
QCD provided one compares the theories at the same values of thermal
masses, rather than equal values of 't Hooft coupling
(although SYM theory has somewhat more soft photons relative to hard
photons in comparison with QCD).  The strongly
coupled theory has a greater photon production rate at large momentum,
relative to the weakly coupled theory,
but less production at small momenta ($k \ll \lambda^{2/3} \,T$).
The production of large mass dilepton pairs is essentially identical
between the weakly and strongly coupled theories.

\section{Coupling \boldmath $\Nfour$ super-Yang-Mills to electromagnetism}
\label{sec:SYM-EM}

The field content of $\Nfour$
SYM theory consists of $SU(\Nc)$ gauge bosons, plus
four Weyl fermions $\psi_p$ and six real%
\footnote{
   It is convenient to regard the scalar fields as components
   of an antisymmetric complex matrix satisfying the
   reality condition
   $(\phi_{pq})^\dagger=\frac12 \, \varepsilon^{pqrs}\phi_{rs}$.
}
scalars $\phi_{pq}\equiv-\phi_{qp}$, $p,q=1,\cdots,4$, transforming in the
adjoint representation of $SU(\Nc)$.
The theory has an anomaly free global $SU(4)$ $R$-symmetry, under which
the fermions transform in the {\bf 4} and the scalars in the ${\bf 6}$.
To model electromagnetic interactions, we add to the theory a
$U(1)$ gauge field coupled to the conserved current corresponding to
a $U(1)$ subgroup of the $SU(4)$ $R$-symmetry.

We will choose the $U(1)$ subgroup
generated by
$t^3\equiv{\rm diag}(\frac12,-\frac12,0,0)$, under which
two of the Weyl fermions have charge $\pm \half$
and two complex scalars have charge $\half$.
The associated conserved current is
\st 
J^{\rm EM}_{\mu }
\equiv
\frac1e \frac{\delta S_{\rm int}}{\delta A^{\mu}}
=
\half \Big[ \psi_1^{a\dagger}\bar\sigma_\mu\psi_1^a
-\psi_2^{a\dagger}\bar\sigma_\mu\psi_2^a
+\sum_{p=3,4} \phi_{1p}^{a\dagger} (-i\vec{D}_\mu 
  +i \!\stackrel{\,_{\leftarrow}}{D}_\mu)\phi_{1p}^a \Big] \,.
\label{eq:Sint}
\stp
A summation over the $SU(\Nc)$ group index $a$ is implied in \Eq{eq:Sint}.
The covariant derivative $D_\mu$ acting on the scalars
involves
both the $SU(\Nc)$ gauge fields and the $U(1)$ electromagnetic potential
$A_\mu$ (with coefficient $\coeff e2$).
However,
the dependence on the $U(1)$ gauge field, reflecting
quadratic dependence on $A_\mu$ in the scalar field part of $S_{\rm int}$,
does not
contribute to the emission rates at leading-order in $e^2$
and can be ignored.%
\footnote{
In the retarded current-current correlator, this term generates
an $O(e^2)$ momentum independent contact term
which does not contribute to the imaginary part of the correlator.
}
So for our purposes we can treat the electromagnetic interaction
as being linear in $A_\mu$, with ${\cal L}_{\rm int} = eJ^3_\mu \, A^\mu$,
where $J^3_\mu$ is the $t^3$ component of the $R$-current in pure
$\Nfour$ SYM.
We further add to the theory one or more ``leptons'' $\ell$
which are fermions with electric charge $e_\ell$,
but with no direct interactions with any SYM fields.
Hence, our complete Lagrange density is
\begin{equation}
  {\cal L} = {\cal L}_{\rm SYM} 
             + {\cal L}_{\rm int}
             - \coeff 14 F_{\mu\nu}^2
             - \bar\ell(\rlap{\,/}D  + m)\ell\,.
\label{eq:SYM-EM}
\end{equation}
We will refer to the theory of \Eq{eq:SYM-EM}
as SYM-EM theory.%
\footnote{
   We would like to stress that, unlike SYM theory, SYM-EM theory
   does not have a known string dual description.
}
For comparisons with QCD,
one should regard
${\cal L}_{\rm SYM}$ as modeling
strongly interacting quark and gluon fields,
while $A_\mu$ describes the photon
and $\ell$ represents the electron and/or muon.

The photons and leptons, once produced, are assumed to stream
through the SYM medium with negligible further interaction,
due to a small value of $e^2$.
Hence their emission rates, to leading-order in $e^2$,
are completely determined by the correlation function
of $R$-currents, $\langle J^3_\mu(0)J^3_\nu(x)\rangle$,
with the expectation value taken in the thermal equilibrium state
of SYM theory.  The evaluation of this correlation function can be
conducted purely within SYM theory, with no further reference to the EM
sector.
Inserting the result into Eqs.~(\ref{eq:photon-rate-general})
and (\ref{eq:dilepton-rate-general}) will yield
the photon and dilepton differential emission rates for $\Nfour$ SYM.

The choice (\ref {eq:Sint}) for the electromagnetic $U(1)$ current 
is not unique.
Choosing an embedding of $U(1)$ within the $SU(4)_R$ symmetry group is
equivalent to choosing a specific linear combination of Cartan subalgebra
generators.
In other words, the most general choice for electromagnetic charge
may be expressed as
\begin{equation}
Q_{\rm EM} = \sum_{a=1}^3 \> \beta_a \, Q^a \,.
\label{eq:embed}
\end{equation}
where
$Q^1 \equiv \frac12 \, {\rm diag}(1,1,-1,-1)$,
$Q^2 \equiv \frac12 \, {\rm diag}(1,-1,1,-1)$, and
$Q^3 \equiv \frac12 \, {\rm diag}(1,-1,-1,1)$
are a convenient set of Cartan generators.
Our particular choice of $U(1)$ embedding, corresponding to $\beta_1 = 0$,
$\beta_2 = \beta_3 = \half$,
gives the charged fermions and charged scalars
equal magnitude charges, and yields an SYM-EM theory which is anomaly free.%
\footnote
    {
    In an arbitrary background $SU(4)$ gauge field,
    the divergence of the $R$-current acquires an anomalous contribution,
    $\partial^\mu J_\mu^a \propto d^{abc} F_{\mu\nu}^b F^{\mu\nu c}$.
    For our chosen $U(1)$ embedding,
    $d^{333}{=}2\,{\rm tr}(t^3 \{t^3,t^3\}){=}0$,
    so our electromagnetic current (\ref {eq:Sint}) is anomaly-free.
    }
It also happens to make the sum of squares of fermion charges
equal in $\Nc=3$, $\Nfour$ SYM and three-flavor QCD.%
\footnote
    {
    In $\Nfour$ SYM this is $\frac{1}{4} (\ncs{-}1)=2$
    (counting Dirac fermion fields),
    while in QCD with real-world charge
    assignments it is $3\times(\frac 49 + \frac 19 + \frac 19)=2$.
    }

The $SU(4)_R$ symmetry of $\Nfour$ SYM guarantees that
the $R$-symmetry current-current correlator
$
    \langle J^a_\mu J^b_\nu \rangle
$
is proportional to $\delta^{ab}$.
[Here $a,b = 1 \cdots 15$ are
$SU(4)$ Lie algebra indices,
and an orthonormal Lie algebra basis is assumed.]
Consequently, if one keeps the choice of $U(1)$ embedding
completely general, as in Eq.~(\ref{eq:embed}),
then the resulting electromagnetic correlator $C^<_{\mu\nu}$
depends on the choice of embedding merely through
an overall normalization factor of $\beta_1^2 + \beta_2^2 + \beta_3^2$
(which is 1/2 for our particular embedding).
Although it would be easy to leave the choice of embedding
completely arbitrary,
to simplify formulas we will use our specific choice
in the next two sections.
In the final discussion we will address the question of what
overall normalization of the electromagnetic current in SYM-EM
is most appropriate when making comparisons with QCD.

\section{Photon and dilepton production rates at strong coupling}
\label{sec:strong-coupling}

The large $N_c$, large 't Hooft coupling limit of $\Nfour$ SYM
theory in a four-dimensional Minkowski space 
at finite temperature $T$ has a dual description in terms of the 
gravitational background with a five-dimensional asymptotically AdS 
metric
\begin{equation}
\label{eq:near-horizon-metric}
ds^2 = 
  \frac{(\pi T R)^2}u
\left[ -f(u) \, dt^2 + dx^2 + dy^2 +dz^2 \right]
 +\frac{R^2}{4 u^2 f(u)} \, du^2\,,
\end{equation}
where $f(u)=1-u^2$, $u\in [0,1]$, 
and $R$ is the curvature radius of the AdS space. 
The metric (\ref{eq:near-horizon-metric}) describes a spacetime with
a horizon at $u = 1$ with Hawking temperature $T$,
and a boundary (where one can regard the dual field theory as residing)
at $u = 0$.

A method for computing the retarded correlation functions
of $R$-currents in the dual gravitational description
was formulated in Refs.~\cite{recipe,hep-th/0205052}.
Subsequently, the locations of singularities of the retarded
correlator $C_{\mu\nu}^{\rm ret}(K)$
in the complex frequency plane were found in Refs.~\cite{NS,quasipaper},
and the spectral function at zero three-momentum
was computed in Ref.~\cite{derek}.
Here we shall determine the spectral function of the $R$-currents
at arbitrary momenta,
and relate them to the photon and dilepton production rates,
respectively.

At zero temperature, the correlation function
has the form dictated by Lorentz and gauge invariance,
\begin{equation}
C_{\mu\nu}^{\rm ret}(K) = P_{\mu\nu}(K) \, \Pi(K^2)\,,
\end{equation}
where $P_{\mu\nu}(K)=\eta_{\mu\nu}{-}K_\mu K_\nu/K^2$
is the usual transverse projector,
and $K^2{\equiv}{-}k_0^2 + \k^2$. For the corresponding spectral function
one finds%
\footnote{
  Except for the overall coefficient, the form of the zero-temperature
  result (\ref{eq:spectral-zero-T}) is completely determined by
  Lorentz, gauge, and scale invariance.
  In zero-temperature SYM, the 
  $R$-current two-point function is protected by non-renormalization theorems,
  and thus is independent of the coupling \cite{Anselmi:1997am}.
  The overall coefficient is therefore fixed by the one-loop
  spectral function evaluated in the free theory.
  In the electromagnetic current (\ref{eq:Sint}),
  the two Weyl fermions can be combined to form one Dirac fermion,
  and for the Feynman correlator one finds
  $\Pi^F(K)=\frac{1}{4\pi}\frac{K^2}{3\pi}\frac{\Nc^2{-}1}{4}
   (1{+}\frac12) [\ln({K^2}/{\mu^2})-2]$.
  Here $\mu$ is the $\overline{\rm MS}$ renormalization scale,
  a factor of $\frac14$ comes from electric charge assignment,
  and the factor of $(1{+}\frac12)$ signifies that there are
  two charged scalars, each contributing one quarter as much as a Dirac
  fermion.  The spectral function is obtained from the relation
  $\Pi^F(K)=\Re \Pi(K) +i\,{\rm sign}(k^0) \Im \Pi(K)$.
  At spacelike momenta, $\Pi^F(K)$ has no imaginary part,
  while at time-like momenta one has to choose $k^0\to k^0{+}i\epsilon$
  which gives $\ln(K^2{-}i\epsilon k^0)/\mu^2 = i\pi +\ln(-K^2)/\mu^2$
  (for positive $k^0$).
  The spectral function $\chi_{\mu\nu}(K)=-2P_{\mu\nu}\Im\Pi(K)$,
  at large $\Nc$,
  is then given by the result (\ref{eq:spectral-zero-T}).
}
\begin{equation}
  \chi_{\mu\nu}(K) =
  -2 \Im C_{\mu\nu}^{\rm ret}(K) =
  P_{\mu\nu}(K) \> \frac{\Nc^2}{16\pi} \> |K^2|\> 
  \Theta(-K^2) \, {\sgn}(k^0)\,.
\label{eq:spectral-zero-T}
\end{equation}
At non-zero temperature, rotation plus gauge invariance implies that the 
correlator has the form
\begin{equation}
   C_{\mu\nu}^{\rm ret}(K) = 
                   P_{\mu\nu}^T(K) \, \Pi^T(k^0,k) +
                   P_{\mu\nu}^L(K) \, \Pi^L(k^0,k)\,,
\end{equation}
where the transverse and the longitudinal projectors are
defined in the standard way as
$P_{00}^T(K)=0$, $P_{0i}^T(K)=0$, $P_{ij}^T(K)=\delta_{ij} - k_i k_j/\k^2$,
and $P_{\mu\nu}^L(K) \equiv P_{\mu\nu}(K) - P_{\mu\nu}^T(K)$.
Spatial indices $i,j$ run over $x,y,z$ and $k \equiv |\k|$.
Thus the trace of the retarded two-point function is
$  \eta^{\mu\nu} C_{\mu\nu}^{\rm ret} = 2\, \Pi^T 
                                + \Pi^L
$
and the trace of the spectral function is
\begin{equation}
 \chimm (k^0,k) = 
   -4\,{\rm Im}\Pi^T(k^0,k) - 2\,{\rm Im}\Pi^L(k^0,k)\,.
\end{equation}
Both $\Pi^T$ and $\Pi^L$ contribute to the dilepton rate,
but only $\Pi^T$ contributes to the photon emission rate, because
the longitudinal part must vanish for lightlike momenta
(otherwise the correlator would be singular on the lightcone).
According to the gauge/gravity duality prescription \cite{AGMOO},
two-point functions of conserved currents in SYM theory
are calculated by analyzing linearized perturbations of a $U(1)$ gauge field 
$A_C$ (having nothing to do with the electromagnetic potential discussed
in the previous section)
on the five-dimensional AdS-Schwarzschild gravitational background
(\ref{eq:near-horizon-metric}).
These perturbations obey Maxwell's equations, 
$\partial_A ( \sqrt{-g} \,g^{AB}g^{CD} F_{BD} ) = 0$,
where $g_{AB}$ is the metric of the background spacetime
(\ref{eq:near-horizon-metric}), and
$F_{BD}=\partial_B A_D - \partial_D A_B$ is the Maxwell field strength.
The Bianchi identity for $F_{BD}$ then implies that
the electric fields $E_i\equiv F_{0i}$
obey the equations \cite{quasipaper}:
\begin{subequations}
\label{eq:electric-fields}
\begin{eqnarray}
    && \Ep'' + \frac{f'}{f} \, \Ep' +
       \frac{\wn^2-\qn^2 f}{u f^2} \, \Ep =0\,, 
\label{eq:Ex}\\
    && E_\parallel '' + \frac{\wn^2 f'}{f (\wn^2 - \qn^2 f)} \, E_\parallel' +
    \frac{\wn^2 - \qn^2 f}{u f^2} \, E_\parallel = 0\,,
\label{eq:Ez}
\end{eqnarray}
\end{subequations}
where $\wn\equiv k^0/(2\pi T)$, $\qn\equiv k/(2\pi T)$,
primes denote derivatives with respect to $u$,
and the subscript refers to the component which is either
perpendicular or parallel to the direction of the three-momentum $\k$.
The equations (\ref{eq:electric-fields}) have singular points at
$u{=}\pm 1, 0$ and $\infty$.%
\footnote
    {
    The longitudinal equation (\ref{eq:Ez}) also has an integrable
    singularity, with exponents 0 and 2,
    at $u^2 = 1-w^2/q^2$.
    When integrating the equation for spacelike Minkowski momenta,
    this singularity may be avoided by making
    an infinitesimal Wick rotation.
    }
At $u{=}1$ (the horizon), the exponents are $\mp i\wn/2$.
These two exponents correspond to two local solutions
representing waves coming into or emerging from the horizon.
To compute the retarded correlators, one has to impose
the incoming wave boundary condition at the horizon,
thus choosing $-i\wn/2$ as the correct exponent \cite{recipe}.
At $u{=}0$ (the boundary), the exponents for both equations
(\ref{eq:electric-fields}) are $0$ and $1$.
Solutions to Eqs.~(\ref{eq:electric-fields})
satisfying the incoming-wave condition at the horizon
can be written as a linear combination of two local solutions near 
the boundary,
\begin{equation}
  E_i(u) = \A_i \, Z^{\rm I}_i(u) + \B_i \, Z^{\rm II}_i(u)\,,
\label{abexp}
\end{equation}
where the index $i$ labels the components of the electric field
(and no summation over $i$ is implied).
The solutions 
$Z^{\rm I}_i$ and $Z^{\rm II}_i$
are given by their standard Frobenius expansions \cite{yellow-book}
near $u{=}0$,
\begin{subequations}
\label{eq:locfrob}
\begin{eqnarray}
  && Z^{\rm I}_i(u) = 1 + h_i Z^{\rm II}_i(u) \ln u + b^{(1)}_{i {\rm I}}\, u 
                   +\cdots\,,\label{locfrob1} \\
  && Z^{\rm II}_i(u) = u\,\big(1 + b^{(1)}_{i{\rm II}} \, u 
                   + b^{(2)}_{i{\rm II}} \, u^2 +\cdots\big)\,.
\label{locfrob2}
\end{eqnarray}
\end{subequations}
All the coefficients $\{ b^{(j)}_{i{\rm I}} \}$
(except $b^{(1)}_{i{\rm I}}$),  $\{ b^{(j)}_{i{\rm II}} \}$,
and $h_i$ are determined by the recursion relations
obtained by substituting the above expansion into the
differential equations (\ref{eq:electric-fields});
for example, $h_i = q^2 - w^2$.
Without loss of generality,
one can set $b^{(1)}_{i{\rm I}}=0$, thus fixing the definition of
$Z^{\rm I}_i(u)$.

The correlators are essentially 
determined by the boundary term of the five-dimensional 
on-shell Maxwell action \cite{recipe,hep-th/0205052,quasipaper}
\begin{equation}
   S_B = \frac{\Nc^2 T^2}{16}
         \lim_{u\to0} \int \frac{ d^4 K}{(2\pi)^4} \left[
         \frac{f}{\qn^2 f{-}\wn^2} \, E_\parallel'(u,K) E_\parallel(u,-K)
         -
         \frac{f}{\wn^2} \, \Ep'(u,K){\cdot}\Ep(u,-K)
         \right]\,.
\end{equation}
Applying the Lorentzian AdS/CFT prescription \cite{recipe}, one finds%
\footnote
    {
    A contact term, proportional to $K^2$,
    is to be discarded in this expression.
    This contact term is real, and does not contribute to the
    physically relevant spectral function.
    }
\begin{equation}
   \Pi^L(k^0,k) = -\frac{\Nc^2 T^2}{8}\lim_{u\to0} 
           \frac{E_\parallel'(u,K)}{E_\parallel(u,K)} \,,\qquad
   \Pi^T(k^0,k) = -\frac{\Nc^2 T^2}{8}\lim_{u\to0} 
           \frac{E_\perp'(u,K)}{E_\perp(u,K)} \,.
\label{eq:PifromE}
\end{equation}
Choosing, for convenience, the three momentum $\k$ to lie
along the $z$ direction,
so that $\Ep =(E_x, E_y)$, and $E_\parallel = E_z$, and using the expansions 
(\ref{abexp}) and (\ref{eq:locfrob}),
the retarded correlation functions reduce to \cite{quasipaper}
\begin{equation}
   \Pi^L(k^0,k) = -\frac{\Nc^2 T^2}{8}  \>
                      \frac{\B_z(k^0,k)}{\A_z(k^0,k)}\,,
  \ \ \ \
   \Pi^T(k^0,k) = -\frac{\Nc^2 T^2}{8} \>
                      \frac{\B_x(k^0,k)}{\A_x(k^0,k)} \,. 
\label{eq:PiBA}
\end{equation}
To evaluate the correlators, one isolates the incoming wave
part of the fluctuating field by finding
solutions to Eq.~(\ref{eq:electric-fields}) of the form%
\footnote{
  The factor $(1{-}u)^{-i\wn/2}$ is dictated by the
  incoming-wave condition at $u{=}1$.
  Separating a factor $(1{+}u)^{-\wn/2}$ in addition is a matter of
  technical convenience.
}
\begin{equation}
  E_i(u) = (1-u)^{-i\wn/2} \> (1+u)^{-\wn/2} \> y_i(u)\,,
\label{eq:def-y}
\end{equation}
where  $y_i(u)$ is regular at $u=1$.
Given the solution [obtained by integrating Eq.~(\ref{eq:electric-fields})
numerically, if necessary],
one may extract the coefficients
$\A_i$ and $\B_i$ from the near-boundary behavior,
and obtain the resulting correlators from Eq.~(\ref{eq:PiBA}).%
\footnote
    {
    Instead of integrating Eq.~(\ref{eq:electric-fields}), with boundary
    condition (\ref{eq:def-y}) outward from the horizon to the boundary,
    and extracting the coefficients $\A_i$ and $\B_i$ from
    the near-boundary behavior,
    improved numerical stability may be obtained if one
    also integrates inward from the boundary to find directly
    the solutions $Z_i^I(u)$ and $Z_i^{II}(u)$ with the prescribed
    boundary behavior (\ref{eq:locfrob}).
    The coefficients $A_i$ and $B_i$ in Eq.~(\ref{abexp})
    may then be determined from the values and derivatives of
    these three solutions at an arbitrary interior point within
    the interval [0,1].
    }

\subsection{Lightlike momenta}
\label{sec:lightlike}
For light-like momenta, $\wn{=}\qn$, inserting the ansatz (\ref{eq:def-y})
into the transverse electric field equation (\ref{eq:Ex})
produces a hypergeometric equation, and yields the analytic solution
\begin{equation}
   E_x(u) =(1-u)^{-i\wn/2} (1+u)^{-\wn/2}\;
 \ofo \Bigl( 1-\half(1{+}i)\wn\,, -\half(1{+}i)\wn\,; 1{-}i\wn; 
\half(1{-}u)\Bigr)\,,
\label{hypergauss}
\end{equation}
where $\ofo (a,b;c;z)$ is Gauss's hypergeometric function.
To extract the imaginary part of the retarded correlator,
which is all we need,
it is convenient to convert (\ref {eq:PifromE}) to the form
\begin{equation}
    \Im \Pi^{T}(k^0,k) = - \frac{\Nc^2 T^2}{8} \,
     \Im \big[ f(u)\, F_\perp(u,K)^*\, F_\perp'(u,K) \big] \,,
\label{eq:Wronskian}
\end{equation}
with
$
 F_\perp(u,K)\equiv { E_x(u,K)}/{ E_x (0,K)}
$.
This formula for $\Im \Pi^T$ reduces to (the imaginary part of)
Eq.~(\ref{eq:PifromE}) in the limit $u \to 0$,
but expression (\ref{eq:Wronskian})
[which is effectively the Wronskian
of $F_\perp$ and its complex conjugate] is actually independent of $u$.
Instead of taking the $u\to 0$ limit,
it is more convenient to evaluate this in the limit $u \to 1$.
Using the solution (\ref{hypergauss}), we find
\begin{equation}
  \Im\, \Pi^T(K) =
  - \frac{\Nc^2 T^2}{8} \, \Im \frac{ i \wn}
{D(\wn)}\,,
\end{equation}
where the denominator is a product of two hypergeometric functions,
\begin{equation}
D(\wn) =  \ofo \Bigl( 1+\half(1{+}i)\wn\,, \half(1{+}i)\wn\,; 1{+}i\wn; 
\half \Bigr) \; \ofo \Bigl( 1-\half(1{+}i)\wn\,, -\half(1{+}i)\wn\,; 1{-}i\wn; 
\half \Bigr)\,.
\end{equation}
With the help of the identity
$
 \ofo \left( a, b; c; z\right) =  (1{-}z)^{-a}\,
  \ofo \left( a, c{-}b; c; {z}/(z{-}1)\right)
$
\cite{abramowitz},
the denominator can be written as 
\begin{equation}
D(\wn) =  4\, \big|\ofo \big(1-\half(1{+}i)\wn,1+\half(1{-}i)\wn;
                  1{-}i\wn;-1\big)\big|^{2}\,.
\end{equation}
Therefore, the spectral function for light-like momenta is
\begin{equation}
   \chimm(k^0{=}k) = - 4 \Im \Pi^T(k^0{=}k) =
   \frac{\Nc^2 T^2 \, \wn}{8}  \>
    \big|\ofo \big(1-\half(1{+}i)\wn,1+\half(1{-}i)\wn;
                  1{-}i\wn;-1\big)\big|^{-2}\,.
\label{eq:trace-photons}
\end{equation}

This result shows that the trace of the spectral function $\chimm(K)$
is manifestly positive, as it should be.
(Note that for light-like momenta, $\chi_{tt}=\chi_{zz}$,
and therefore $\chimm=2\chi_{xx}$.)
The asymptotic behavior for
small and large frequencies is%
\footnote
    {
    These asymptotics are derived in appendix~\ref{app:photons-asympt}.
    A simple approximation which is asymptotically correct
    and accurate to better than 2\% for all frequencies is
    $
	\chimm(\wn{=}\qn)
	\approx
	\half\ncs T^2 \wn
	\big(
	    1 + \coeff 1{729}
	    [\coeff{\sqrt\pi}{2} \, 3^{5/6}
	    \Gamma(\coeff 23)/\Gamma(\coeff 13)]^{24}
	    \wn^4
	\big)^{1/24}
	/(1+\coeff 13 \pi^2 \wn^2)^{1/4}
    $.
    }
\begin{equation}
  \chimm(\wn{=}\qn) \sim \begin{cases}  
\half{\Nc^2 T^2} \, \big[ \wn {-} \frac{\pi^2}{12}\, \wn^3 + O(\wn^5)\big]\,,
  & \wn \ll 1\,; \\
& \\
 \coeff 14 {\Nc^2 T^2}\, \wn^{2/3} \,
3^{5/6} \, {\Gamma(\coeff 23)}/{\Gamma(\coeff 13)} + O(1)\,, & \wn \gg 1\,.
\end{cases}
   \label{f_grav}
\end{equation}

\begin{figure}
  \psfrag{chi}{\hspace{-0.7cm}$\chimm(\qn{=}\wn)/\wn$}
  \psfrag{w}{\hspace{-0.8cm}$\wn\equiv k^0/(2\pi T)$}
  \includegraphics[width=3.5in]{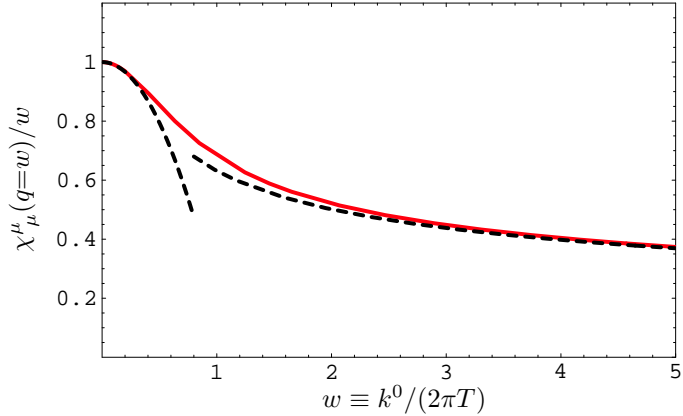}
\caption{
	Trace of the spectral function for lightlike momenta
	divided by frequency, 
	$\eta^{\mu\nu}\chi_{\mu\nu}(\wn{=}\qn)/\wn$,
	in units of $\half\Nc^2 T^2$, plotted as a function of frequency,
	with $\wn\equiv k^0/(2\pi T)$ and $q \equiv |\k|/(2\pi T)$.
	At small frequency,
	$\chimm(\wn{=}\qn)/\wn$ approaches a constant limiting value,
	while at large frequency $\chimm(\wn{=}\qn)/\wn$
	falls as $\wn^{-1/3}$.
	The solid (red) line shows the exact result (\ref{eq:trace-photons})
	while the dashed lines show the low- and high-frequency
	asymptotics (\ref{f_grav}).
}
\label{fig:gb-factor}
\end{figure}

A graph of the trace of the spectral function at light-like momentum,
together with the asymptotics (\ref{f_grav}),
 is shown in Fig.~\ref{fig:gb-factor}.
The leading small-frequency behavior agrees with that found earlier in
Ref.~\cite{hep-th/0205052}, where it was used to evaluate the
diffusion constant of $R$-charge in SYM theory.
The expression (\ref{eq:trace-photons}) for the spectral function is valid
to leading order in the limit of large $\nc$ and large 
't~Hooft coupling.
This result shows that the photon production rate for $\Nfour$ SYM theory
approaches a finite limit as $\lambda\to\infty$.

\subsection{Timelike and spacelike momenta}
\label{sec:timelike}

\begin{figure}[h]
  \begin{minipage}[t]{.48\textwidth}
  \psfrag{kmin}{\footnotesize $q_-$}
  \psfrag{kplus}{\footnotesize \quad$q_+$}
  \psfrag{chiT}{\hspace{-1.6cm}
                $\Delta\chi^T(\wn,\qn)$}
  \includegraphics[width=3.0in]{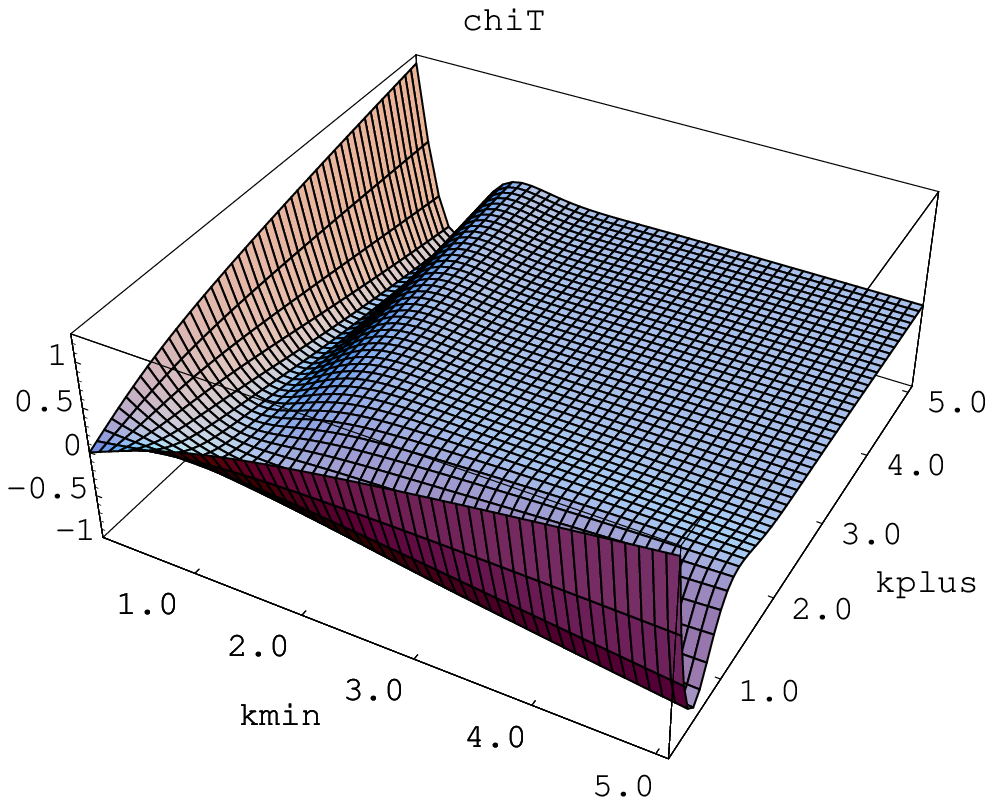}
  \end{minipage}
  \begin{minipage}[t]{.48\textwidth}
  \psfrag{kmin}{\footnotesize $q_-$}
  \psfrag{kplus}{\footnotesize \quad$q_+$}
  \psfrag{chiL}{\hspace{-1.6cm}
                $\Delta\chi^L(\wn,\qn)$}
  \includegraphics[width=3.0in]{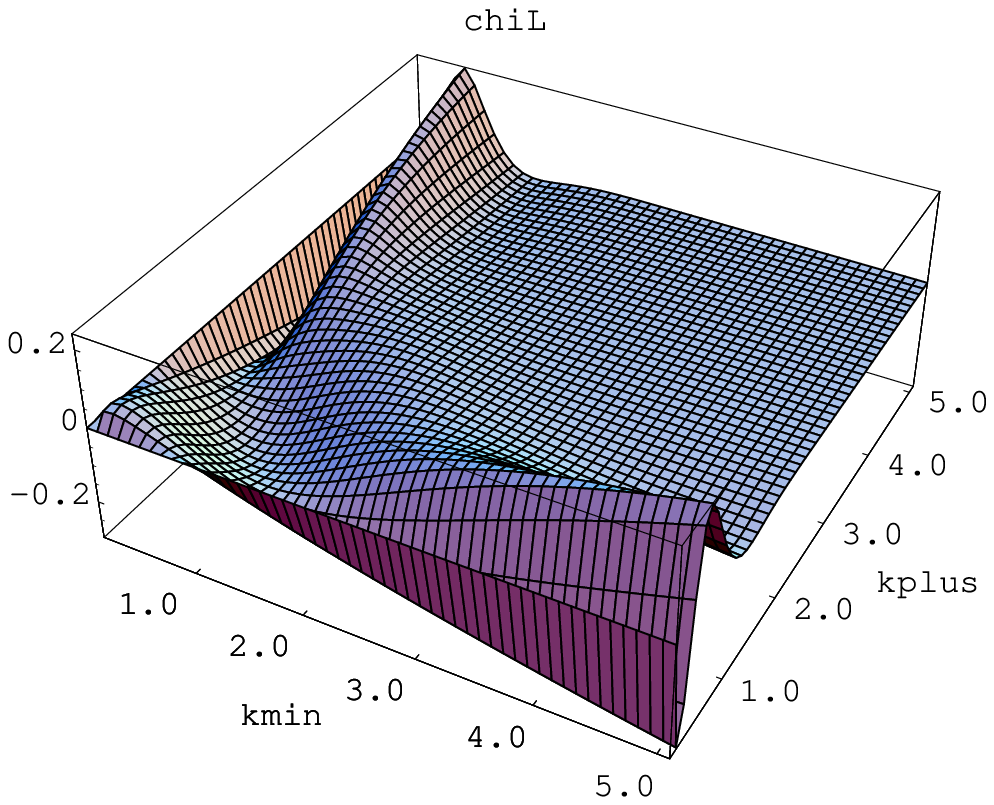}
  \end{minipage}
  \caption{
  Transverse and longitudinal
  spectral functions for time-like momenta,
  shown in the $(\wn,\qn)$ plane.
  Axes are $q_\pm = \wn\pm\qn$; the tip of the light-cone
  is on the left.
  The graphs show finite-temperature contributions to
  $\chi^T\equiv\chi_{xx}+\chi_{yy}$ (left), and
  $\chi^L\equiv-\chi_{tt}+\chi_{zz}$ (right),
  plotted in units of $\frac12\Nc^2 T^2$.
  The subtracted zero-temperature contributions are
  $\chi^T(\wn,\qn)|_{T{=}0} = \pi(\wn^2{-}\qn^2)$, and
  $\chi^L(\wn,\qn)|_{T{=}0} = \frac{\pi}{2}(\wn^2{-}\qn^2)$.
  Note that $\chi^L(\wn,\qn)$ is zero on the light-cone
  because $\chi_{tt}(k^0{=}k)=\chi_{zz}(k^0{=}k)$.
}
\label{fig:chi-timelike-3d}
\end{figure}

\begin{figure}
  \hspace*{-10pt}
\psfrag{w}{\raisebox{-1ex}{\footnotesize\hspace{-0.8cm}$w = k^0/(2\pi T)$}}
\psfrag{chimm}{$\chimm(\wn,\qn)$}
\psfrag{chimmw}{$\chimm(\wn,\qn)/\wn$}
  \includegraphics[width=3.3in]{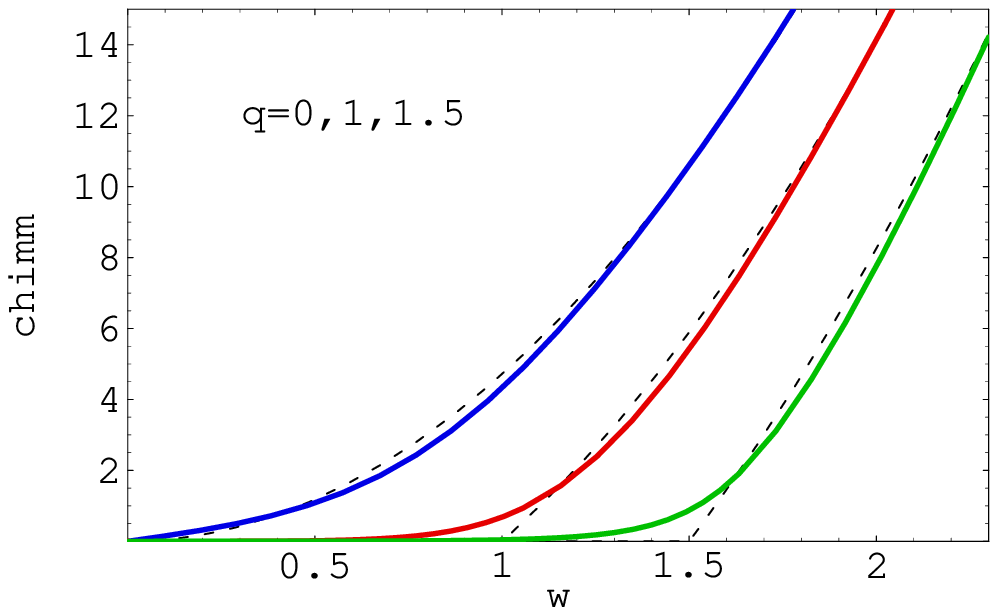}
  \includegraphics[width=3.3in]{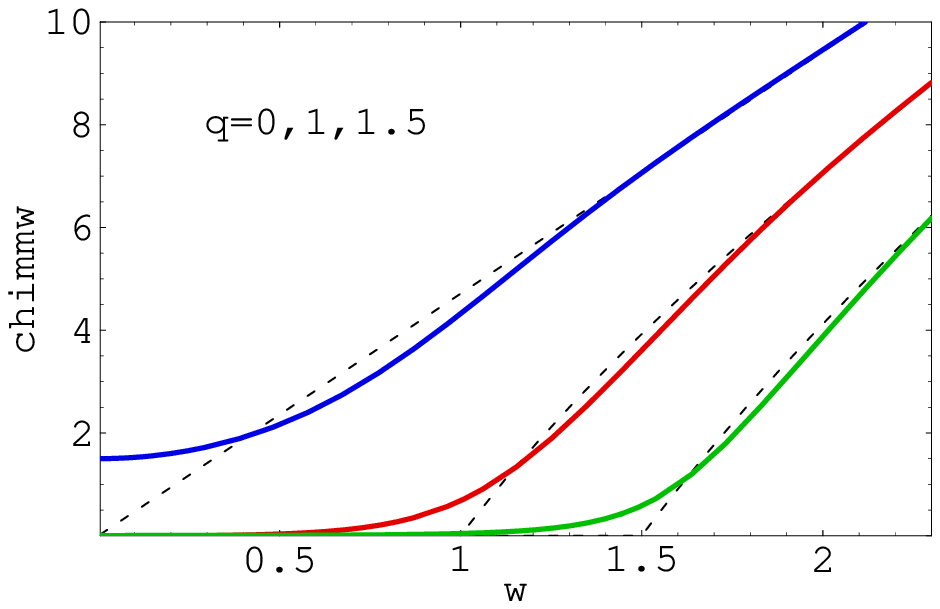}%
  \hspace*{-10pt}
  \caption{
  Spectral function trace $\chimm(k^0,k)$ (left)
  and $\chimm(k^0,k)/w$ (right),
  in units of $\Nc^2 T^2/2$,
  plotted as a function of $w \equiv k^0/(2\pi T)$.
  The different curves correspond to differing values of the momentum;
  from left to right,
  $q \equiv k/(2\pi T) = 0, 1.0, 1.5$.
  The dotted black lines show the zero-temperature result.
}
\label{fig:chimm-vs-w}
\end{figure}

\begin{figure}
\psfrag{w}{\raisebox{-1ex}{\footnotesize\hspace{-0.8cm}$w = k^0/(2\pi T)$}}
\psfrag{delChimm}{$\Delta\chimm(\wn,\qn)$}
  \includegraphics[width=3.8in]{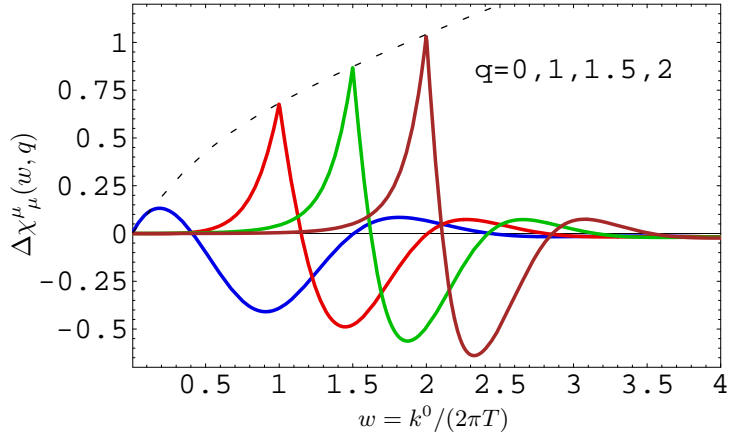}
  \caption{
  Deviation of the spectral function trace $\chimm(k^0,k)$
  from its zero temperature limit,
  in units of $\Nc^2 T^2/2$,
  as a function of $w \equiv k^0/(2\pi T)$.
  The different curves correspond to differing values of the momentum;
  $q \equiv k/(2\pi T) = 0$ (blue), 1.0 (red), 1.5 (green), and 2.0 (brown).
  The curves at non-zero momentum have cusps on the light cone
  [at $w=1, 1.5$, and 2, respectively]
  where the zero-temperature result first turns on in a non-analytic fashion.
  The dotted line showing the envelope of the cusps is the plot
  of $\chimm$ on the lightcone.
}
\label{fig:delta-chimm-vs-w}
\end{figure}

At time-like momenta, both $\Pi^T$ and $\Pi^L$ contribute
to $\chimm(k^0,k)$.
The mode equations (\ref{eq:electric-fields}) cannot be solved
analytically for arbitrary frequency and wavevector, so we determine
the spectral function numerically, as explained above.
The result is shown in Fig.~\ref{fig:chi-timelike-3d},
where we plot the temperature-dependent portion of the
transverse and longitudinal contributions to the
spectral function in the $(k^0,k)$ plane,
defined as
\begin{eqnarray}
    \Delta\chi^T(k^0,k) &\equiv& 
    P^T_{\mu\nu}(K)
    \left[ \chi^{\mu\nu}(k^0,k) - \chi^{\mu\nu}_{T=0}(k^0,k) \right]
\nonumber\\&=&
    \chi_{xx}(k^0,k)+\chi_{yy}(k^0,k) - \frac{N_c^2}{8\pi}
    \left[(k^0)^2{-}k^2 \right]
    \Theta((k^0)^2{-}k^2) \, \sgn(k^0) \,,
\\
    \Delta\chi^L(k^0,k) &\equiv& 
    P^L_{\mu\nu}(K)
    \left[ \chi^{\mu\nu}(k^0,k) - \chi^{\mu\nu}_{T=0}(k^0,k) \right]
\nonumber\\&=&
    \chi_{zz}(k^0,k)-\chi_{tt}(k^0,k) - \frac{N_c^2}{16\pi}
    \left[(k^0)^2{-}k^2 \right]
    \Theta((k^0)^2{-}k^2) \, \sgn(k^0) \,.
\end{eqnarray}
The complete result for $\chimm$ is plotted in Fig.~\ref{fig:chimm-vs-w}
as a function of frequency for several values of the spatial momentum.
As these plots show,
$\chimm$ rapidly approaches the zero-temperature curve
as the frequency increases.
The oscillatory finite-temperature deviations are shown directly
in Fig.~\ref{fig:delta-chimm-vs-w}.

\begin{figure}
  \hspace*{-20pt}
\psfrag{q}{\raisebox{-1ex}{\footnotesize\hspace{-0.8cm}$q = k/(2\pi T)$}}
\psfrag{chiT}{$\chi^T(\wn,\qn)$}
\psfrag{chiL}{$\chi^L(\wn,\qn)$}
  \includegraphics[width=3.3in]{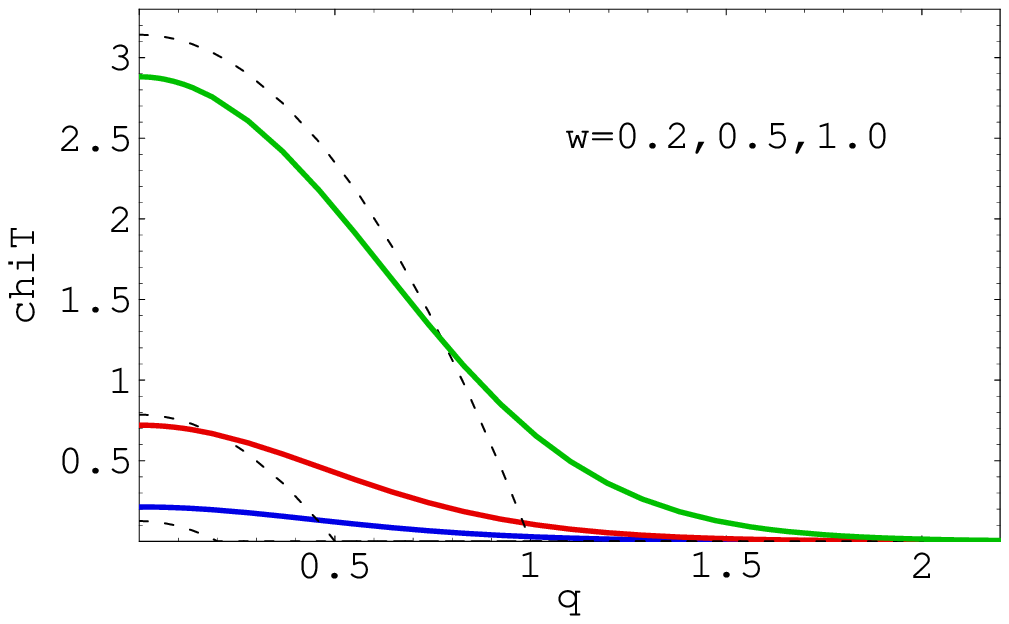}
  \kern 5pt
  \includegraphics[width=3.3in]{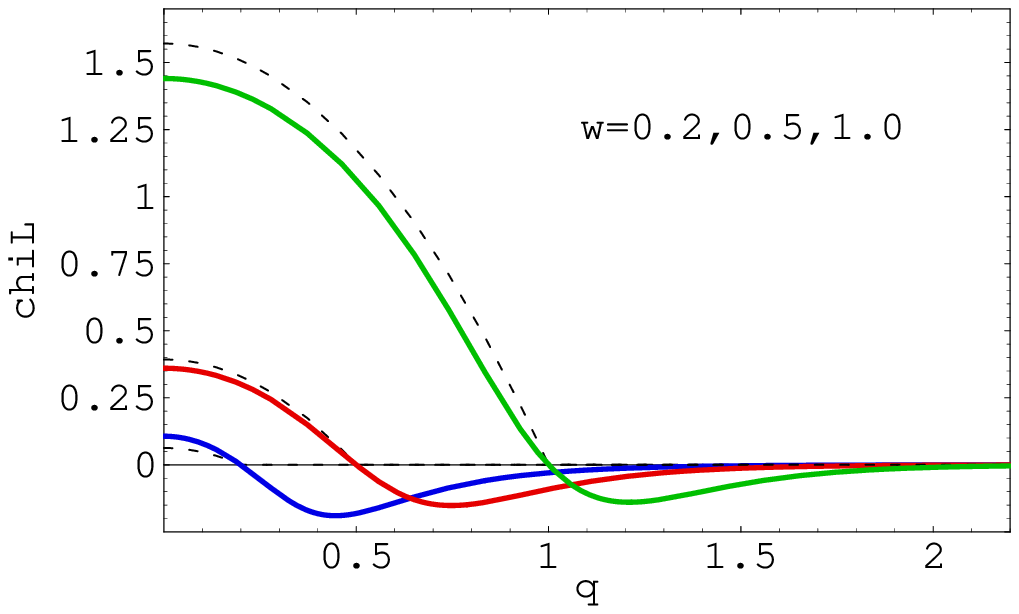}
  \hspace*{-15pt}
  \caption{
  Transverse (left) and longitudinal (right) spectral functions plotted
  as a function of $q = k/(2\pi T)$, for several values of the
  frequency, in units of $\ncs T^2/2$.
  The different curves correspond to differing values of the frequency;
  from left to right,
  $w \equiv k^0/(2\pi T) = 0.2$ (blue), 0.5 (red), and 1.0 (green).
  The dashed black lines show the corresponding zero temperature result.
}
\label{fig:chiTL-vs-q}
\end{figure}

Slices of transverse and longitudinal spectral densities at fixed frequency,
plotted as a function of spatial momentum,
are shown in Fig.~\ref{fig:chiTL-vs-q}.
Note that the longitudinal spectral density is not always positive
(for positive frequencies).
The components $\chi_{zz}$ and $\chi_{tt}$ are individually
both positive (for positive frequencies), but their difference
$\chi_L = \chi_{zz} - \chi_{tt}$ can have either sign.
As one moves deeper into the spacelike region, the spectral densities
rapidly decrease.

\begin{figure}[t]
  \hspace*{-20pt}
\psfrag{w}{\raisebox{-1ex}{\footnotesize\hspace{-0.8cm}$w = k^0/(2\pi T)$}}
\psfrag{chittw}{\raisebox{0.5ex}{$\chi_{tt}(\wn,\qn)/\wn$}}
\psfrag{chizzw2}{\raisebox{0.5ex}{$\chi_{zz}(\wn,\qn)/\wn^2$}}
  \includegraphics[width=3.4in]{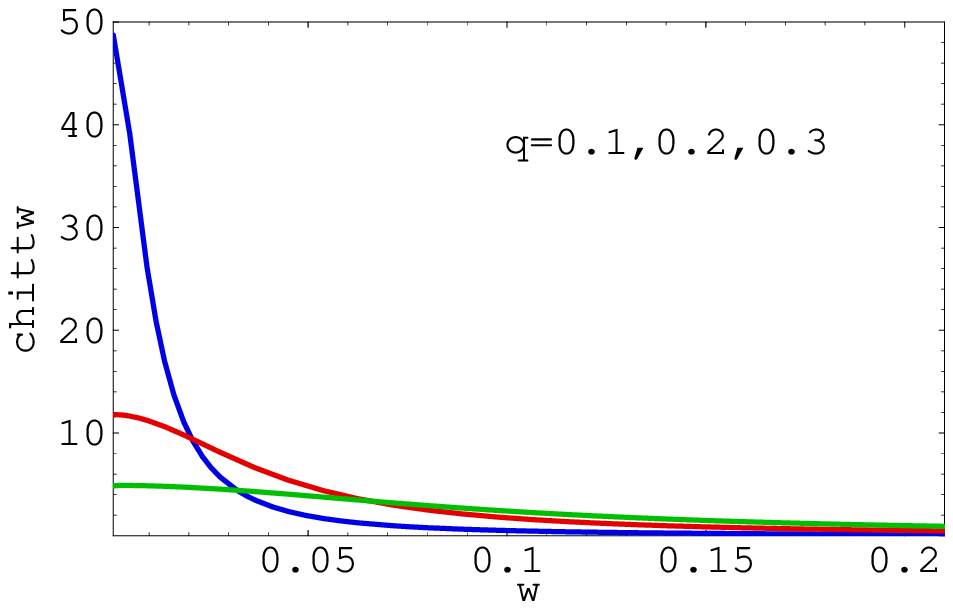}
  \hspace*{-10pt}
  \includegraphics[width=3.4in]{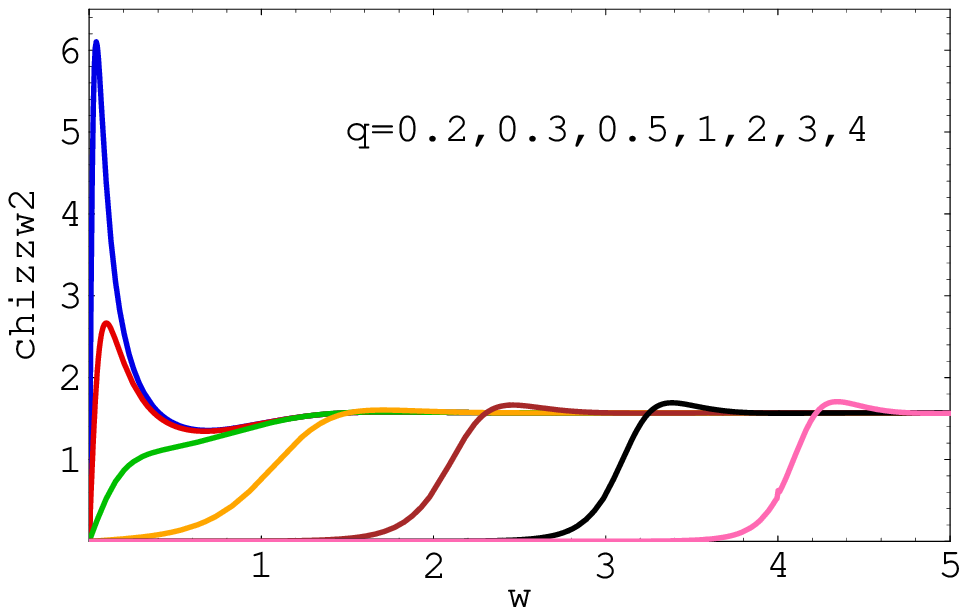}%
  \hspace*{-20pt}
  \vspace*{-10pt}
  \caption{
  Left:
  Time-time spectral density divided by frequency,
  $\chi_{tt}(k^0,k)/w$,
  plotted as a function of $w \equiv k^0/(2\pi T)$, for
  $q \equiv k/(2\pi T) = 0.1$ (blue), 0.2 (red), and 0.3 (green).
  In the limit of vanishing spatial momentum,
  $\chi_{tt}/w$ approaches a delta-function
  in frequency.
  Right:
  Longitudinal space-space spectral density
  divided by frequency squared,
  $\chi_{zz}(k^0,k) /w^2$
  as a function of frequency, for
  $q \equiv k/(2\pi T) = 0.2$ (blue), 0.3 (red), 0.5 (green), 1 (orange),
  2 (brown), 3 (black) and 4 (pink).
  One sees the diffusive peak for small frequency and momentum,
  together with the approach to the zero-temperature result
  at higher frequency.
}
\label{fig:chitt-chizz}
\end{figure}

At small frequency and small momentum,
the longitudinal spectral density $\chi_L = \chi_{zz} - \chi_{tt}$
has hydrodynamic structure which cannot be resolved in
Figs.~\ref{fig:chimm-vs-w} and \ref{fig:chiTL-vs-q}.
The time-time and longitudinal space-space components should behave as
\begin{equation}
    \chi_{tt}(\omega,k)
    \sim
    \frac {2 \omega D k^2}{\omega^2 + (D k^2)^2} \> \Xi \,,\qquad
    \chi_{zz}(\omega,k)
    \sim
    \frac {2 \omega^3 D}{\omega^2 + (D k^2)^2} \> \Xi \,,
\label{eq:hydro-forms}
\end{equation}
where $\omega\equiv k^0$, $D$ is the $R$-charge diffusion constant, and
$\Xi \equiv \beta \langle Q^2 \rangle / (\mbox{volume})$
is the charge susceptibility.
As the spatial momentum $k \to 0$, $\chi_{tt}(\omega,k)/\omega$
approaches a delta-function in frequency (times $2\pi\Xi$).
This behavior is shown on the left in Fig.~\ref{fig:chitt-chizz}.
The longitudinal space-space spectral function $\chi_{zz}$,
divided by $w^2$, is displayed on the right in Fig.~\ref{fig:chitt-chizz}
as a function of frequency for various values of momentum.
The Ward identity for the correlator implies that
$\chi_{zz}(\omega,k) = \frac {\omega^2}{k^2} \, \chi_{tt}(\omega,k)$,
so $\chi_{tt}$ and $\chi_{zz}$ contain exactly the same information.
But with this scaling, 
one sees both the diffusive hydrodynamic peak at small frequency
for the low momentum curves,
together with the approach of all curves to a common
high frequency value of $\coeff \pi 4 \, \ncs T^2$.
This constant value is precisely the zero temperature result
for $\chi_{zz}(\omega,k)/w^2$.
Our results are consistent, as they must be, with
the value of the diffusion constant
previously found in Ref.~\cite{hep-th/0205052},
\begin{equation}
    D = \frac 1{2\pi T} \,.
\end{equation}
One may also easily extract the charge susceptibility
of strongly coupled SYM,%
\footnote{
  Ref.~\cite{hep-th/0205052} found that
  $\chi_{xx}(\omega,k)=(\Nc^2 T/8\pi) \, \omega$,
  and
  $\chi_{tt}(\omega,k)=(\Nc^2 T/8\pi) \, \omega k^2/[\omega^2+(Dk^2)^2]$
  with $D=1/2\pi T$.
  Comparison with the form (\ref{eq:hydro-forms}) immediately gives
  the stated value of the susceptibility which is,
  of course, consistent with the Kubo formula
  $D\Xi=\lim_{\omega\to0} \frac 1{6}\, \omega^{-1} \chimm(\omega,k{=}0)$.
}
\begin{equation}
    \Xi = \coeff 18 \ncs T^2 \,.
\label{eq:Xi}
\end{equation}

\begin{figure}[ht]
\psfrag{qplus}{\raisebox{-1ex}
  {\footnotesize\hspace{-0.8cm}$q^+ = (k^0+k^3)/(2\pi T)$}}
\psfrag{chimm}{$\chimm(q^+,q^-)$}
  \includegraphics[width=4.0in]{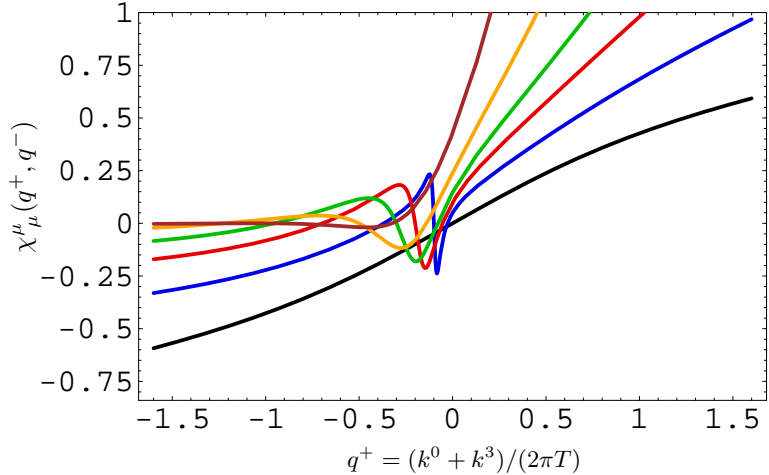}%
  \caption{
  Spectral density trace as a function of
  $q^+ \equiv (k^0+k^3)/(2\pi T)$
  for various values of $q^- \equiv (k^0 - k^3)/(2\pi T)$,
  plotted in units of $\ncs T^2/2$.
  The left side of the plot, $q^+ < 0$ corresponds to spacelike momenta,
  while the right side, $q^+ > 0$, is the timelike region.
  The various curves (from bottom to top, except near the origin)
  correspond to
  $q^- = 0$ (black), 0.1 (blue), 0.2 (red), and 0.3 (green),
  0.5 (orange), and 1.0 (brown).
}
\label{fig:chimm-vs-q+}
\end{figure}

Finally, in Fig.~\ref{fig:chimm-vs-q+} we plot $\chimm$ as a function
of $q^+ \equiv (k^0 {+} k^3)/(2\pi T)$ for various values of
$q^- \equiv (k^0 {-} k^3)/(2\pi T)$.
The $q^-=0$ curve corresponds to light-like momenta;
this curve is an odd function of $q^+$.
As one moves away from the lightcone by increasing $q^-$,
the curves with $q^-$ small compared to 1 clearly show
hydrodynamic ``wiggles'' at small $q^+$, but this structure
broadens and becomes washed out at larger values of $q^-$.

\newpage

\subsection{Electrical conductivity}

The electrical conductivity $\sigma$ can be computed by using the Kubo formula
(\ref{eq:Kubo-sigma})
expressing the conductivity in terms of the zero-frequency limit
of $\eta^{\mu\nu} C^<_{\mu\nu}(K)$, or equivalently the zero-frequency slope
of the trace of the spectral function of electromagnetic currents
(times $T$).
The small-frequency behavior of the spectral function
of $R$-currents in strongly coupled SYM theory was analyzed
in Ref.~\cite{hep-th/0205052}.
Inserting the limiting low frequency behavior
found in that work into
the Kubo formula (\ref{eq:Kubo-sigma}),
gives
\begin{equation}
  \sigma = e^2 \, \frac{\Nc^2 \, T}{16\pi} \,.
\label{eq:sigma}
\end{equation}
Using the low frequency behavior of the spectral density (\ref{f_grav})
for null momenta to evaluate the lightlike
Kubo formula (\ref{eq:Kubo2})
yields the same value, as it must.

The result (\ref {eq:sigma}) demonstrates that the conductivity is finite
and coupling-independent in the limit of large coupling.
Note that $\sigma$ is sensitive to the total number of
degrees of freedom in the theory, and therefore is not
directly useful as a means of comparing transport properties
in different theories.
A more ``universal'' quantity is obtained by dividing the conductivity
by the charge susceptibility (\ref {eq:Xi}), giving
\begin{equation}
  \frac{\sigma}{e^2 \Xi} = \frac{1}{2\pi T} \,,
\end{equation}
in strongly coupled SYM.
This is precisely the diffusion constant of $R$-charge \cite{hep-th/0205052},
showing the consistency of the Einstein relation $\sigma/(e^2\Xi)=D$.

\subsection{Thermal resonances?}

In a confining theory like QCD, the spectral density of the
zero-temperature current-current correlator will have delta-function
contributions from mesons like the $\rho$ and $J/\psi$,
plus narrow peaks from other hadronic resonances
(with widths vanishing as $\Nc\to\infty$).
The delta functions will acquire thermal widths (which are also $1/\nc$
suppressed)
at non-zero temperature,
but for some range of temperatures one will see easily
recognizable resonances with widths small compared to their energies.
There is some evidence from lattice studies that
the $J/\psi$ remains a well-defined resonance even at temperatures
of a few times $T_{\rm c}$
\cite{Jpsi-lattice1,Jpsi-lattice2,Jpsi-lattice3,Jpsi-lattice4},
and there has been recent discussion of possible signs
of other ``bound states'' above $T_{\rm c}$ \cite{Shuryak}.

Since it is a conformal theory,
$\Nfour$ SYM has no particle spectrum
and the zero temperature current-current spectral density
(\ref{eq:spectral-zero-T})
is featureless.
However, one may mock up a confining theory by considering deformations
of the gravitational description of $\Nfour$ SYM in which one cuts off
the AdS space at some value of $u = u_c$,
so the coordinate $u$ ranges from
$u=0$ (the boundary) to $u=u_c$ (the cutoff).
According to AdS/CFT duality,
string theory on this cut-off geometry
should describe a large $\Nc$ field theory with a mass gap,
and hence a discrete spectrum of bound states,
determined by the eigenvalues of the corresponding
wave equations.%
\footnote{
  Such ``hard-wall'' cut-off models were discussed from the earliest days of
  AdS/CFT, and represent the simplest version of how confinement
  may be realized in the dual gravity description \cite{wall1,wall2,wall3}.
}
When the field theory is considered
at non-zero temperature,
there will be a confinement/deconfinement transition
at a (non-zero) critical temperature $T_c$.
For temperatures below $T_c$, the relevant dual
gravitational geometry remains the same as at zero temperature
(but with time periodically identified when analytically continued
to Euclidean signature).
Above $T_c$, the field theory will be in a deconfined plasma phase,
and the appropriate dual geometry is
AdS-Schwarzschild (times $S^5$).
In the simple hard-wall model,
properly comparing the gravitational action of these geometries
(which determines the free energy of the thermal field theory)
shows that the transition to the AdS-Schwarzschild
geometry occurs when the hard-wall cutoff is inside the horizon
\cite{chris}.

In this model, the spectral function of $R$-currents
in the low-temperature phase
is temperature independent, and equal to a sum
of discrete delta functions,
whose locations are determined by
the energies of the bound states,
which are eigenvalues of normalizable fluctuations in the cut-off
AdS geometry.
Above the critical temperature, the hard-wall cutoff is hidden
by the horizon, and is entirely irrelevant to physics outside the horizon.
The spectral function of $R$-currents is precisely the same as in
pure $\Nfour$ SYM.
The results of section~\ref{sec:timelike}
explicitly show that the spectral functions have
no structure which could be interpreted as
peaks corresponding to narrow resonances which survive
in the high-temperature phase.
Thus in the hard-wall model, bound states ``dissolve''
completely at the confinement/deconfinement transition.
Whether this reflects physical features which may be shared
by real QCD,
or is just a pathology of the hard-wall AdS/CFT model,
is not completely clear.
We suspect it is a generic feature of light hadrons in confining gauge
theories at $\nc=\infty$.

\section{Photon and dilepton production rates at weak coupling}
\label{sec:weak-coupling}

\subsection{Dilepton production}

\begin{figure}
\centerbox{0.6}{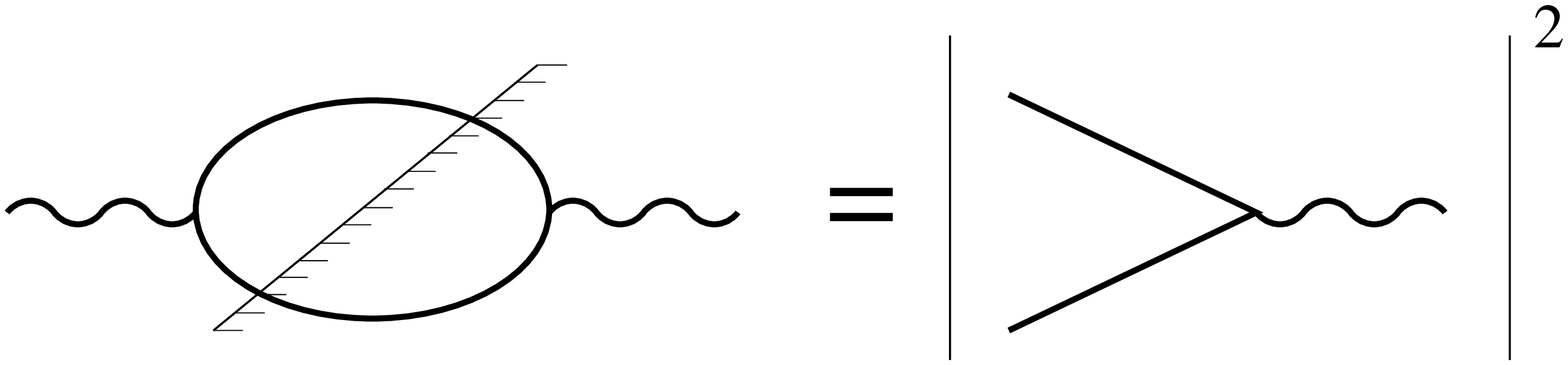}
\caption{\label{fig_dilepton_graphs}
Graph needed to compute the dilepton production rate to lowest order.
The diagonal slash represents a ``cut'' through the diagram;
the solid line is either a fermion or a scalar.}
\end{figure}

When $\lambda$ is sufficiently small,
one may use weak-coupling methods to compute
the photon and dilepton emission rates.
The easiest process to analyze perturbatively is the dilepton production
rate, because it arises already at $O(\lambda^0)$.  The simplest way to
structure the calculation is to compute $\eta^{\mu\nu}C^<_{\mu\nu}(K)$
directly.
This requires evaluating the single cut, one loop graph shown in
Figure \ref{fig_dilepton_graphs}.  The cut lines have the propagator
replaced by the appropriate statistical function times the discontinuity
in the propagator (the difference between $+i\epsilon$ and
$-i\epsilon$ prescriptions), which at this order means the substitution
of $-i/(p^2{+}m^2)$ by $2\pi \, n(p^0) \, \delta(p^2{+}m^2)$.
Noting that the sum of the charge squared for all SYM Weyl fermions
coincides with that for the charged scalars,
and equals $\half(\ncs-1)$,
the leading order result for the Wightman function is%
\footnote
    {
    The weak-coupling results of this section
    are valid for arbitrary $\nc$.
    }
\begin{eqnarray}
\eta^{\mu\nu} C^<_{\mu\nu}(K) & = & -\half(\ncs{-}1)\int \frac{d^4 P}{(2\pi)^4} \>
2\pi\delta(P^2) \> 2\pi \delta((K{-}P)^2)
  \Big\{
  \nbose(p^0) \, \nbose(k^0{-}p^0) \>
  (2P{-}K)^2
\nonumber \\ && \hspace{1.2in} {}
+ 
\nfermi(p^0) \, \nfermi(k^0{-}p^0)  \>
  \Tr \big[{\half (1{-}\gamma^5)}
       \nott{P} \gamma^\mu (\nott{K}{-}\nott{P}) \gamma_\mu \big]
  \Big\}
\nonumber \\
& = & -\half(\ncs{-}1)\int \frac{p^2 dp \> d\cos\theta_{pk}}{4\pi p} \>
    \delta(K^2 + 2pk^0 -2pk\cos\theta_{pk}) 
\nonumber \\ && \hspace{1.2in} {}
    \times K^2 \, \Big\{ 
  \nbose(p^0) \, \nbose(k^0{-}p^0)
  +2 
  \nfermi(p^0) \, \nfermi(k^0{-}p^0)
\Big\} \,,
\end{eqnarray}
where we used $(P{-}K)^2=0$ to set $2P\cdot K=K^2$.
The integrals are straightforward and give
\begin{eqnarray}
\eta^{\mu\nu} C^<_{\mu\nu}(K) & = & \frac{(\ncs{-}1) }{16\pi} \frac{(-K^2)}{k}
   \int_{\frac{k^0{-}k}{2}}^{\frac{k^0{+}k}{2}} dp \left[
   2\nfermi(p)\,\nfermi(k^0{-}p) + \nbose(p)\,\nbose(k^0{-}p) \right]
\nonumber \\
& = & \frac{(\ncs{-}1)}{16\pi} \, (-K^2) \, \nbose(k^0) \!\left[ 3 
   - \frac{2T}{k} \ln \frac{1{+}e^{-(k^0-k)/{2T}}}{1{+}e^{-(k^0+k)/{2T}}}
   + \frac{T}{k} \ln \frac{1{-}e^{-(k^0+k)/{2T}}}{1{-}e^{-(k^0-k)/{2T}}}
   \right] \!,
\label{eq:C-LO}
\end{eqnarray}
for timelike $K$, $K^2<0$.
Inserting this result
into Eq.~(\ref{eq:dilepton-rate-general}) yields the actual
dilepton emission rate.
To express this in terms of the spectral weight, one must merely remove
the factor $\nbose(k^0)$; the result differs from the vacuum result by the two
logarithmic factors inside the square bracket, which vanish
exponentially for $(k^0-k)$ large compared to $T$.

\subsection{Photon production from \boldmath $2 \leftrightarrow 2$ scattering}

The one-loop result (\ref{eq:C-LO}) for $\eta^{\mu\nu} C^<_{\mu\nu}(K)$,
which is independent of $\lambda=g^2 \nc$,
vanishes on the light cone, $K^2 = 0$.
Consequently, the spectral weight for lightlike momenta first
arises at the two-loop level.  Physically, the timelike spectral weight
represents the splitting of a timelike virtual photon into a pair of charged
particles, a process which occurs even in the absence of strong interactions.
In contrast, the lightlike spectral weight represents real photon
production, which only occurs via scattering processes and therefore
involves powers of $\lambda$ and higher loop orders.  A complication is
that in a thermal system, the expansion of physical quantities in powers
of $\lambda$ is {\em not\/} the same as a diagrammatic expansion in the
number of loops.
This is a consequence of sensitivity to energy and momentum scales
which are parametrically small compared to $T$.
For lightlike momentum, this complication arises at
the first nontrivial order, and requires an infinite resummation
of diagrams to find the leading order weak-coupling photon production rate
\cite{AMY2}.  This rate can be understood as the sum of a contribution
from Compton-like $2\leftrightarrow 2$ scattering processes
\cite{Baier,Kapusta} and near-collinear bremsstrahlung
and pair-annihilation processes
\cite{Aurenche}, which are further corrected due to the LPM effect
\cite{LP,M1,M2}.  A complete treatment for the QCD plasma is given in
Refs.~\cite{AMY3,AMY2}, and we will extend it here to the case at hand.
The main new complications are the appearance of scalar fields and Yukawa
couplings in the $2\leftrightarrow 2$ processes, and the addition of
bremsstrahlung from charged scalars, which fortunately was already
discussed in Ref.~\cite{AMY2}.

\begin{figure}[t]
\centerbox{0.65}{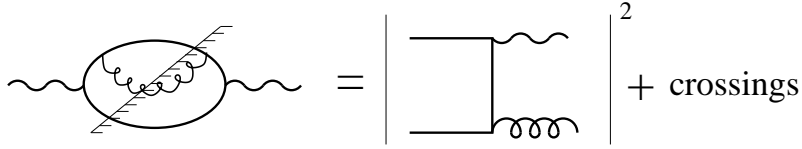}
\caption{\label{fig_22_diagram} Relation of the 2-loop contribution
to the current-current 
correlator and tree-level $2\leftrightarrow 2$ scattering diagrams.
Interference contributions arise when the gluon runs from one solid line
to the other (not shown).}
\end{figure}

Consider first the $2\leftrightarrow 2$ particle processes where two SYM
excitations collide to produce an SYM excitation and a photon.  They
arise in the current-current correlator at the two loop level, as
illustrated in Figure \ref{fig_22_diagram}.  Calculation of these
contributions involves
integrating the squared matrix element for each
possible production process over all possible momenta of the SYM
particles, with appropriate population functions and Pauli blocking or
Bose stimulation functions on final states.
The resulting contribution to the photon production rate has the form
\begin{equation}
e^2 \eta^{\mu\nu} C^<_{\mu\nu,2\leftrightarrow 2}(K) 
  = \int \frac{d^3 \p \, d^3 \p' \, d^3 \k'}
{(2\pi)^9 2p^0 2p'{}^0 2k'{}^0} \> (2\pi)^4 \delta^4(P{+}P'{-}K{-}K')
\sum_{acd} |{\cal M}^{ac}_{\gamma d}|^2 \,
n_a(p) n_c(p') [1{\pm} n_d(k')] \, ,
\label{rate_22}
\end{equation}
where $\p,\p'$ represent the momenta of incoming particles of type $a,c$,
$\k'$ is the momentum of an outgoing particle of type $d$,
$n_a=\nfermi$ or $\nbose$ according to the statistics of species $a$, and
the $\pm$ sign is + if $d$ is a boson and $-$ if $d$ is a fermion.
(Note that $[1{\pm} n_d(k')]=e^{k'/T} n_d(k')$ in either case.)  All external
states can be treated as massless, since thermal corrections to their
dispersion relations are suppressed by a power of $\lambda$;
therefore $p^0\equiv p=|\p|$.  The sum $\sum_{acd}$ runs over species type,
color, spin (including the photon
spin), and particle/antiparticle where appropriate.  We have computed
these summed matrix elements for the SYM theory under consideration; the
result, organized by the spins of external
states, is presented in Table \ref{table_Msq}.  

\begin{table}
\begin{tabular}{|c|c|c|}\hline
Process & Diagrams & $\sum |{\cal M}|^2 = e^2 \lambda (\ncs{-}1) \times {}$ \\ \hline
$FF \rightarrow \gamma G$ &
\begin{picture}(150,35)
\put(10,-10){\begin{picture}(75,25)
  \thicklines
  \put(0,0){\line(1,0){30}}
  \put(0,30){\line(1,0){30}}
  \put(30,0){\line(0,1){30}}
  \multiput(30,0)(0,30){2}{\circle*{3}}
  \multiput(36,0)(8,0){3}{\oval(12,12)[t]}
  \multiput(40,0)(8,0){2}{\oval(4,4)[b]}
  \multiput(32,30)(8,0){4}{\oval(4,4)[t]}
  \multiput(36,30)(8,0){3}{\oval(4,4)[b]}
  \end{picture}}
\put(80,-10){\begin{picture}(75,35)
  \thicklines
  \put(0,0){\line(1,1){30}}
  \put(0,30){\line(1,-1){30}}
  \put(30,0){\line(0,1){30}}
  \multiput(30,0)(0,30){2}{\circle*{3}}
  \multiput(36,0)(8,0){3}{\oval(12,12)[t]}
  \multiput(40,0)(8,0){2}{\oval(4,4)[b]}
  \multiput(32,30)(8,0){4}{\oval(4,4)[t]}
  \multiput(36,30)(8,0){3}{\oval(4,4)[b]}
  \end{picture}}
\end{picture} &
    ${\displaystyle 2 \left( \frac ut + \frac tu \right)   }$ \\
$FG \rightarrow \gamma F$ &
\begin{picture}(150,35)
\put(10,-10){\begin{picture}(75,35)
  \thicklines
  \put(30,0){\line(1,0){30}}
  \put(0,30){\line(1,0){30}}
  \put(30,0){\line(0,1){30}}
  \multiput(30,0)(0,30){2}{\circle*{3}}
  \multiput(8,0)(8,0){3}{\oval(12,12)[b]}
  \multiput(12,0)(8,0){2}{\oval(4,4)[t]}
  \multiput(32,30)(8,0){4}{\oval(4,4)[t]}
  \multiput(36,30)(8,0){3}{\oval(4,4)[b]}
  \end{picture}}
\put(80,-10){\begin{picture}(75,35)
  \thicklines
  \put(30,30){\line(1,-1){30}}
  \put(0,30){\line(1,-1){30}}
  \put(30,0){\line(0,1){30}}
  \multiput(30,0)(0,30){2}{\circle*{3}}
  \multiput(8,0)(8,0){3}{\oval(12,12)[b]}
  \multiput(12,0)(8,0){2}{\oval(4,4)[t]}
  \multiput(32,30)(8,0){4}{\oval(4,4)[t]}
  \multiput(36,30)(8,0){3}{\oval(4,4)[b]}
  \end{picture}}
\end{picture} &
    ${\displaystyle 4 \left( -\frac ts - \frac st \right) }$ \\
$FF \rightarrow \gamma S$ &
\begin{picture}(220,35)
\put(10,-10){\begin{picture}(75,35)
  \thicklines
  \put(0,0){\line(1,0){30}}
  \put(0,30){\line(1,0){30}}
  \put(30,0){\line(0,1){30}}
  \multiput(30,0)(0,30){2}{\circle*{3}}
  \multiput(30,0)(4,0){8}{\circle*{2}}
  \multiput(32,30)(8,0){4}{\oval(4,4)[t]}
  \multiput(36,30)(8,0){3}{\oval(4,4)[b]}
  \end{picture}}
\put(80,-10){\begin{picture}(75,35)
  \thicklines
  \put(0,0){\line(1,1){30}}
  \put(0,30){\line(1,-1){30}}
  \put(30,0){\line(0,1){30}}
  \multiput(30,0)(0,30){2}{\circle*{3}}
  \multiput(30,0)(4,0){8}{\circle*{2}}
  \multiput(32,30)(8,0){4}{\oval(4,4)[t]}
  \multiput(36,30)(8,0){3}{\oval(4,4)[b]}
  \end{picture}}
\put(150,-10){\begin{picture}(75,35)
  \thicklines
  \put(0,0){\line(1,0){30}}
  \put(0,30){\line(1,-1){30}}
  \multiput(30,0)(0,30){2}{\circle*{3}}
  \multiput(30,30)(3,-3){11}{\circle*{2}}
  \multiput(30,0)(0,4){8}{\circle*{2}}
  \multiput(32,30)(8,0){4}{\oval(4,4)[t]}
  \multiput(36,30)(8,0){3}{\oval(4,4)[b]}
  \end{picture}}
\end{picture} &
    ${\displaystyle 2 \left( 3\,\frac ut + 3\,\frac tu + 2 \right) }$ \\
$FS \rightarrow \gamma F$ &
\begin{picture}(220,35)
\put(10,-10){\begin{picture}(75,35)
  \thicklines
  \put(30,0){\line(1,0){30}}
  \put(0,30){\line(1,0){30}}
  \put(30,0){\line(0,1){30}}
  \multiput(30,0)(0,30){2}{\circle*{3}}
  \multiput(0,0)(4,0){8}{\circle*{2}}
  \multiput(32,30)(8,0){4}{\oval(4,4)[t]}
  \multiput(36,30)(8,0){3}{\oval(4,4)[b]}
  \end{picture}}
\put(80,-10){\begin{picture}(75,35)
  \thicklines
  \put(30,30){\line(1,-1){30}}
  \put(0,30){\line(1,-1){30}}
  \put(30,0){\line(0,1){30}}
  \multiput(30,0)(0,30){2}{\circle*{3}}
  \multiput(00,0)(4,0){8}{\circle*{2}}
  \multiput(32,30)(8,0){4}{\oval(4,4)[t]}
  \multiput(36,30)(8,0){3}{\oval(4,4)[b]}
  \end{picture}}
\put(150,-10){\begin{picture}(75,35)
  \thicklines
  \put(30,0){\line(1,0){30}}
  \put(0,30){\line(1,-1){30}}
  \multiput(30,0)(0,30){2}{\circle*{3}}
  \multiput(30,30)(-3,-3){11}{\circle*{2}}
  \multiput(30,0)(0,4){8}{\circle*{2}}
  \multiput(32,30)(8,0){4}{\oval(4,4)[t]}
  \multiput(36,30)(8,0){3}{\oval(4,4)[b]}
  \end{picture}}
\end{picture} &
    $\quad {\displaystyle 4 \left( -3\, \frac ts -3 \, \frac st -2 \right) 
    }\quad$ \\
$\quad SS \rightarrow \gamma G \quad$ &
\begin{picture}(220,35)
\put(10,-10){\begin{picture}(75,35)
  \thicklines
  \multiput(0,0)(4,0){8}{\circle*{2}}
  \multiput(0,30)(4,0){8}{\circle*{2}}
  \multiput(30,0)(0,4){8}{\circle*{2}}
  \multiput(30,0)(0,30){2}{\circle*{3}}
  \multiput(36,0)(8,0){3}{\oval(12,12)[t]}
  \multiput(40,0)(8,0){2}{\oval(4,4)[b]}
  \multiput(32,30)(8,0){4}{\oval(4,4)[t]}
  \multiput(36,30)(8,0){3}{\oval(4,4)[b]}
  \end{picture}}
\put(80,-10){\begin{picture}(75,35)
  \thicklines
  \multiput(0,0)(3,3){11}{\circle*{2}}
  \multiput(0,30)(3,-3){11}{\circle*{2}}
  \multiput(30,0)(0,4){8}{\circle*{2}}
  \multiput(30,0)(0,30){2}{\circle*{3}}
  \multiput(36,0)(8,0){3}{\oval(12,12)[t]}
  \multiput(40,0)(8,0){2}{\oval(4,4)[b]}
  \multiput(32,30)(8,0){4}{\oval(4,4)[t]}
  \multiput(36,30)(8,0){3}{\oval(4,4)[b]}
  \end{picture}}
\put(165,-10){\begin{picture}(75,35)
  \thicklines
  \multiput(0,0)(3,3){6}{\circle*{2}}
  \multiput(0,30)(3,-3){6}{\circle*{2}}
  \put(15,15){\circle*{4}}
  \multiput(15,9)(4,-4){4}{\oval(12,12)[tr]}
  \multiput(19,9)(4,-4){3}{\oval(4,4)[b]}
  \multiput(19,9)(4,-4){3}{\oval(4,4)[tl]}
  \multiput(17,15)(4,4){4}{\oval(4,4)[tl]}
  \multiput(17,19)(4,4){4}{\oval(4,4)[br]}
  \end{picture}}
\end{picture} &
    $4$ \\
$SG \rightarrow \gamma S$ &
\begin{picture}(220,42)
\put(10,-10){\begin{picture}(75,35)
  \thicklines
  \multiput(30,0)(4,0){8}{\circle*{2}}
  \multiput(0,30)(4,0){8}{\circle*{2}}
  \multiput(30,0)(0,4){8}{\circle*{2}}
  \multiput(30,0)(0,30){2}{\circle*{3}}
  \multiput(8,0)(8,0){3}{\oval(12,12)[t]}
  \multiput(12,0)(8,0){2}{\oval(4,4)[b]}
  \multiput(32,30)(8,0){4}{\oval(4,4)[t]}
  \multiput(36,30)(8,0){3}{\oval(4,4)[b]}
  \end{picture}}
\put(80,-10){\begin{picture}(75,35)
  \thicklines
  \multiput(30,30)(3,-3){11}{\circle*{2}}
  \multiput(0,30)(3,-3){11}{\circle*{2}}
  \multiput(30,0)(0,4){8}{\circle*{2}}
  \multiput(30,0)(0,30){2}{\circle*{3}}
  \multiput(8,0)(8,0){3}{\oval(12,12)[b]}
  \multiput(12,0)(8,0){2}{\oval(4,4)[t]}
  \multiput(32,30)(8,0){4}{\oval(4,4)[t]}
  \multiput(36,30)(8,0){3}{\oval(4,4)[b]}
  \end{picture}}
\put(165,-10){\begin{picture}(75,35)
  \thicklines
  \multiput(30,0)(-3,3){6}{\circle*{2}}
  \multiput(0,30)(3,-3){6}{\circle*{2}}
  \put(15,15){\circle*{4}}
  \multiput(15,9)(-4,-4){4}{\oval(12,12)[tl]}
  \multiput(11,9)(-4,-4){3}{\oval(4,4)[b]}
  \multiput(11,9)(-4,-4){3}{\oval(4,4)[tr]}
  \multiput(17,15)(4,4){4}{\oval(4,4)[tl]}
  \multiput(17,19)(4,4){4}{\oval(4,4)[br]}
  \end{picture}}
\end{picture} &
    $8$ \\ &\begin{picture}(10,15)\end{picture}&\\ \hline
\end{tabular}
\caption{\label{table_Msq} Fully summed squared matrix elements for all
  processes, organized by the spin of the participants:  $F$=spin-1/2
  fermion, $S$=spin-0 scalar, and $G$=spin-1 gluon.  The summation over
  spin, species label, color, and particle/antiparticle has already been
  conducted; for instance, the $FF\rightarrow \gamma S$ contribution
  includes production of both neutral and charged scalars, of either
  charge.  For the processes involving both fermion and scalar lines,
  which two of the three diagrams contribute depends on which two of the
  three external states carry electric charge.}
\end{table}

Those matrix elements with $1/t$ or $1/u$ behavior lead to small-angle
divergences in the photon production rate.  The best way to see this is to
choose coordinates with the $z$ axis aligned with $\k$.  One shifts
integration variables from $\p$ to $\q\equiv \k-\p$ and uses the spatial
momentum conserving $\delta$-function to perform the $\k'$ integration.
Introducing a dummy integration variable $\omega$ via
\begin{equation}
1 = \int d\omega \; \delta(\omega + k^0 - p^0) \, ,
\end{equation}
the integration measure in \Eq{rate_22} can be reduced to \cite{AMY3}
\begin{equation}
\int \frac{d^3 \p \, d^3 \p' \, d^3 \k'}
{(2\pi)^9 2p^0 2p'{}^0 2k'{}^0} \> (2\pi)^4 \delta^4(P{+}P'{-}K{-}K')
= \frac{1}{(4\pi)^3 k} \int_0^\infty dq
\int_{-q}^{{\rm min}[q,2k{-}q]} d\omega
\int_{\frac{q{+}\omega}{2}}^\infty dp' \int_0^{2\pi} \frac{d\phi}{2\pi}
\, ,
\label{eq:integration}
\end{equation}
and in terms of these variables, $u/t \simeq 2kp'(1{-}\cos\phi)/q^2$ at
small $q$.
(In this section, $q{\equiv}|\q|$, not to be confused with
the normalized momentum in section \ref{sec:strong-coupling}.)
The $1/q^2$ behavior of the squared matrix element makes up for the
two powers of $q$ in
the $dq$ and $d\omega$ integrations, leading to a logarithmically divergent result.
Of course, the photon production rate is not actually divergent; for
sufficiently small $q^2$, the calculation of the matrix element
presented so far is insufficient and requires plasma corrections to the
internal propagator which is responsible for the $1/t$ (or $1/u$) behavior.
These corrections become important when $q^2 \sim \lambda T^2$, and are
referred to as Hard Thermal Loop (HTL) corrections \cite{Braaten}.  The
correction moderates the small $q$ behavior of the matrix element and
renders the production rate finite, albeit with an extra logarithmic dependence
on $1/\lambda$.

The coefficient of the log is quite easy to compute, using the above
behavior of $u/t$ and the quoted matrix elements for various processes.
We find it by extracting the small $q$ behavior of \Eq{rate_22}
and applying it to the region $\lambda T^2 \ll q^2 \ll T^2$.
The resulting small $q$ contribution to $\eta^{\mu\nu} C^<_{\mu\nu}$ is
\begin{eqnarray}
\eta^{\mu\nu} C^<_{\mu\nu,2\leftrightarrow 2}(K)
 & \simeq & 32 \, \frac{\lambda (\ncs{-}1)}{(4\pi)^3 k}
\int_{q_{\rm min}}^{q_{\rm max}} \frac{dq}{q} \int_{-q}^{q} \frac{d\omega}{q}
\int_{0}^\infty dp' \int_0^{2\pi} \frac{d\phi}{2\pi} \>
\nfermi(k) \,
\frac{2p'k\,  (1{-}\cos\phi)\, e^{p'/T}}{(e^{p'/T}{+}1)(e^{p'/T}{-}1)}
\nonumber \\
& \simeq & \frac{\lambda (\ncs{-}1) T^2 \, \nfermi(k)}{4\pi} \, \ln \left(
\frac{q_{\rm max} \sim T}{q_{\rm min} \sim T\sqrt{\lambda}} \right) \,.
\end{eqnarray}
Therefore, the log-enhanced part of the photon production
rate is
\begin{equation}
\eta^{\mu\nu} C^<_{\mu\nu}(K) = \frac{\lambda (\ncs{-}1) T^2 \, \nfermi(k)}{4\pi}
\left[ \ln \lambda^{-1/2} + O(1) \right] \, .
\end{equation}
We will use this coefficient to normalize all other contributions to the
photon production rate, which can be written as
\begin{equation}
\eta^{\mu\nu} C^<_{\mu\nu}(K) = \frac{\lambda (\ncs{-}1) T^2 \, \nfermi(k)}{4\pi}
\left[ \ln \lambda^{-1/2} + C_{\rm tot}(k/T) + O\big(\sqrt\lambda\big) \right] \, .
\label{eq:Ctot}
\end{equation}

Some, but not all,
contributions to $C_{\rm tot}$ may be extracted from
\Eq{rate_22}.  This is done by evaluating carefully the $q^2\sim T^2$
region (using the full matrix element and phase space) and the $q^2 \sim
\lambda T^2$ region (using small $q^2$ approximations but including HTL
corrections in the matrix element).  The small $q^2$ region has already
been handled in the literature \cite{Baier,Kapusta},%
\footnote{%
    It is not obvious that the previous analysis \cite{Baier,Kapusta},
    which treated the soft momentum region in ordinary QCD, can be
    applied unmodified to ${\cal N}{=}4$ SYM.  They can, however.  First,
    note that it is only processes involving fermions (quarks) which
    give rise to the log, which arises when the quark momentum is
    small.  Second, the result in the literature only depended on the
    form of the quark self-energy at soft momentum---the fermionic hard
    thermal loop (HTL) self-energy.  This is actually the same, up to
    the overall coefficient, between QCD and SYM, even though 
    3/4 of the SYM fermionic self-energy comes about from
    interactions with the scalars.  To see this, recall that
    the gluon contribution to the HTL fermion self-energy
    arises from the loop integration (in Feynman gauge),
    \begin{equation}
    \Sigma(Q) = \sumint_K \gamma^\mu S(Q{+}K) \gamma^\nu G_{\mu\nu}(K)
    = \sumint_K \eta_{\mu\nu} \gamma^\mu (\nott{Q}{+}\nott{K}) \gamma^\nu
    \frac{1}{(K+Q)^2} \frac{1}{K^2} \,.
    \end{equation}
    The gauge choice is irrelevant when we take the $(Q \ll K)$ HTL piece,
    since
    this piece is gauge invariant.  The Yukawa interactions give rise to
    a loop integral of form
    \begin{equation}
    \Sigma(Q) = -2 \, \sumint_K S(Q{+}K) \Delta(K) =
    -2 \, \sumint_K (\nott{Q}{+}\nott{K}) \,
    \frac{1}{(K+Q)^2} \frac{1}{K^2} \, .
    \end{equation}
    Using $\gamma^\mu \gamma^\alpha \gamma_\mu =
    -2\gamma^\alpha$, the gluonic loop contribution immediately
    collapses to the same result.}
and the hard region can be handled by numerical quadratures integration
of \Eq{rate_22} after using \Eq{eq:integration} to reduce it to a triple
integral.  A similar set of integration variables is available for the
constant and $t/s$ type matrix elements, see Ref.~\cite{AMY3}.  For large
$k/T$, the coefficient behaves as $\half \ln(2k/T)$ plus a constant.
It is convenient to separate this asymptotic behavior
so we will write this contribution to $C_{\rm tot}(k/T)$ as 
$C_{2\leftrightarrow 2}(k/T)+\half\ln(2k/T)$.  Results for this quantity are
given below in subsection~D.

\subsection{Near-collinear bremsstrahlung and pair annihilation}

\begin{figure}[h]
\centerbox{0.5}{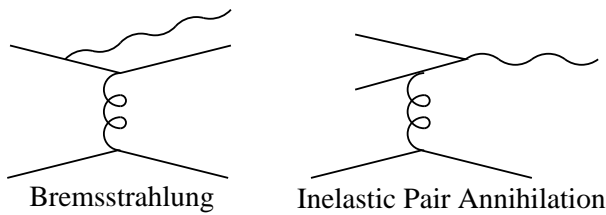}
\caption{\label{fig:brem} Basic processes behind bremsstrahlung and
inelastic pair annihilation.  The gluon exchange leads to a small angle
Coulombic scattering, and the photon is produced by nearly collinear
initial or final state radiation.}
\end{figure}

Besides the $2\leftrightarrow 2$ processes just considered, photons are also
produced at leading order by bremsstrahlung and
inelastic pair annihilation, illustrated in Figure \ref{fig:brem}.  These
contributions arise because small-angle Coulombic
scattering is very efficient;  
the rate per particle of Coulomb scattering in the thermal
medium is $O(\lambda T)$, rather than $O(\lambda^2 T)$ as might
naively be expected.  And as usual in gauge theories, initial or final
state radiation is an efficient process; photon radiation occurs in
$O(e^2)$ of such scatterings.  This leads to a photon production rate
which is $O(\lambda e^2 T^4)$, the same order as the $2\leftrightarrow
2$ processes just considered.%
\footnote
    {
    Analogous processes involving scalar exchange do not have the
    same soft enhancement, and hence are subleading.
    }

Unfortunately, the calculation of photon production via these processes
is a little more complicated than just evaluating the graphs of
Fig.~\ref{fig:brem}.  The physical reason that
initial and final state radiation is so efficient is that the wavefunction
of the radiated particle, emerging at a small (collinear) angle,
can overlap with the emitter for a long time, so the amplitude
builds up coherently over this
large formation time.  But in a medium, further scatterings may occur
within this coherence time; photon radiation from different scattering
events can be partially coherent, as noted by Landau over 50 years ago
\cite{LP,M1,M2}.  One should therefore consider emission from a charge carrier
as it moves through the medium, making a series of small-angle
scatterings.  The photon emission vertex can appear at any point along
the trajectory; in computing the probability for an emission, one must
integrate over this time separately in the
amplitude for the process and the conjugate of the amplitude.
Hence, there is an integral over the time difference between the
photon vertex in the amplitude and the conjugate amplitude.  Because the
energy of a state with a particle of momentum $(p{+}k)$ differs from the
energy of a state with a particle of momentum $p$ and a photon of
momentum $k$, there is a phase difference which grows as the time difference
becomes large.
One must correctly incorporate the effects of multiple small-angle
scatterings occurring during this extended emission process.
The photon production rate is then determined by
summing the resulting photon production from a particular charged
particle of momentum $p{+}k$ over all charges in the medium.  Leaving
the detailed derivation to references \cite{AMY2,AMY3}, the
contribution from these processes to the current-current correlator boils
down to
\begin{eqnarray}
\eta^{\mu\nu} C^<_{\mu\nu,{\rm brem}+{\rm pair}}(K) &=& 
    \frac{\ncs{-}1}{2}
    \int_{-\infty}^\infty \frac{dp}{2\pi} \left(
    \frac{ 
    \nfermi(k{+}p)\,[1{-}\nfermi(p)] 
    \big[ p^2+(p{+}k)^2 \big]}{4p^2(p{+}k)^2}
    + \frac{ 
    \nbose(k{+}p)\,[1{+}\nbose(p)]}{2p(p{+}k)}
    \right) 
\nonumber \\
&& \hspace{1.2in} \times
\int \frac{d^2 \p_\perp}{(2\pi)^2}\>
\Re \big[ 2\p_\perp \cdot \f(\p_\perp,p,k) \big]
\, ,
\label{eq:finalint}
\end{eqnarray}
where the function $\f(\p_\perp,p,k)$ is the solution to the
linear integral equation
\begin{eqnarray}
2\p_\perp &=& 
    \frac{i k [\p_\perp^2{+}m_\infty^2]}{2p(k{+}p)} \, \f(\p_\perp,p,k) +
    \int d^2 \q_\perp \>
    \frac{d\Gamma_{\rm scatt}}{d^2 \q_\perp} \,
    \big[ \f(\p_\perp,p,k) - \f(\p_\perp{+}\q_\perp,p,k) \big] 
\, .
\label{integral_eq}
\end{eqnarray}
In the final integral (\ref{eq:finalint}),
$p{+}k$ is the initial state energy of the particle radiating the
photon and $p$ is its final state energy; when $p<0$ the process is pair
annihilation (note that $[1\pm n(p)]=n(-p)$ so the final state
blocking/stimulation function becomes an initial state population function) 
and when $(p{+}k)<0$ the antiparticle is the initial particle.  
The difference in coefficients between the fermion and scalar
contributions in the integral (\ref{eq:finalint})
reflects their different DGLAP kernels for photon emission.
The integral equation (\ref{integral_eq}) accounts for the
evolution of the mixed state $|p{+}k\rangle \langle p,k|$ through the
plasma, that is, for the evolution after photon emission in the
amplitude but before photon emission in the conjugate amplitude.  It has
been Fourier transformed into frequency, which makes it easier to
evaluate but harder to interpret; the $2\p_\perp$ comes from the dot product
of the photon polarization tensor with the current; the first, imaginary term
accounts for the phase due to the energy difference, the second term
accounts for scattering events.
In the imaginary term, $m_\infty^2$ is the dispersion correction that a large
momentum $(p \gg \sqrt\lambda T)$ particle receives due to the thermal medium,
$p_{\rm on{-}shell}^0 \simeq p+m_\infty^2/2p$.
This turns out to be identical for scalar, spinor,
and gauge degrees of freedom in $\Nfour$ SYM,
\begin{equation}
    m_\infty^2 = \lambda T^2 \,.
\label{eq:minfty}
\end{equation}
Gelis {\it et al.} \cite{Gelis}
derived a very compact
expression for the differential cross-section for scattering with
transverse momentum exchange $\q_\perp$ (after integrating over the
longitudinal momentum exchange),
\begin{equation}
\frac{(2\pi)^2 d\Gamma_{\rm scatt}}{d^2 \q_\perp} = \lambda T \,\frac{\mD^2}
{\q_\perp^2(\q_\perp^2+\mD^2)} \, ,
\end{equation}
where $\mD^2 = 2m_\infty^2 = 2\lambda T^2$ is the static Debye screening mass.

\subsection{Photon production results}

The integral equation (\ref{integral_eq})
can be solved by variational methods or by Fourier
transformation into a differential equation.  The same equation appears
for both scalar and fermionic contributions because the scalars
and fermions have the same small-angle cross-section and the same
dispersion correction (\ref{eq:minfty}).  The resulting contributions,
normalized to the leading-log coefficient of \Eq{eq:Ctot},
are presented separately in Table \ref{table:results}
as the coefficients $C_{\rm pair}$
(from the region $-k<p<0$) and $C_{\rm brem}$ (from $p>0$ and $p<-k$),
in addition to the combined value
\begin{equation}
C_{\rm tot}(k/T) =
\half \ln \frac{2k}{T} + C_{\rm 2\leftrightarrow 2}(k/T)
 + C_{\rm brem}(k/T) + C_{\rm pair}(k/T) \, .
\label{eq:Atot}
\end{equation}
Our numerical results, within the range $0.2 < k/T < 20$,
are reproduced quite accurately by the approximate forms
\begin{eqnarray}
    C_{2\leftrightarrow2}(x)
    &\simeq&
    2.01 \, x^{-1} - 0.158 - 0.615 \, e^{-0.187 \, x} \,,
\label {eq:approxC2} 
\\ \noalign{\hbox{and}}
    C_{\rm brem}(x) + C_{\rm pair}(x)
    &\simeq&
    0.954 \, {x^{-3/2}} \, \ln(2.36 + 1/x) + 0.069 + {0.0289 \, x}  \,.
\label {eq:approxC}
\end{eqnarray}
The fitting form for $C_{2\leftrightarrow 2}$ has absolute accuracy
of 0.02 in this range, and the form for $C_{\rm brem}+C_{\rm pair}$ has
relative accuracy better than $2\%$.
Inserting the result for $C_{\rm tot}(k/T)$ into the leading order
form (\ref{eq:Ctot}) for the correlator, and then multiplying by
photon phase space as shown in Eq.~(\ref{eq:photon-rate-general}), yields
the actual photon emission rate.
This is plotted in the next Section.

\def\sp{$\;\;$}
\def\z{\phantom 0}
\def\o{\phantom 1}
\def\m{\phantom { $-$}}
\begin{table}
\begin{tabular}{@{\extracolsep{10pt}}|c|cccc|} \hline
~~~$k/T$~~~& $C_{\rm brem}$ & $C_{\rm pair}$ & $C_{\rm 2\leftrightarrow 2}$
  & $C_{\rm tot}$ \\ \hline
\z0.10 & 69.9040\o & 1.32650 &   19.318681\o & 89.7444\z \\
\z0.15 & 34.0596\o & 0.886328 &   12.650618\o & 46.9946\z \\
\z0.20 & 20.3471\o & 0.666836 &    9.315910 & 29.8717\z \\
\z0.30 & 9.77708  & 0.448540 &    5.980593 & 15.9508\o \\
\z0.40 & 5.79022  & 0.340596 &    4.312890 & 10.3321\o \\
\z0.50 & 3.85278  & 0.276800 &    3.312841 & 7.44243 \\
\z0.75 & 1.84384  & 0.194596 &    1.983390 & 4.22456 \\
\o1.0 & 1.10442  & 0.156616 &    1.325792 & 2.93340 \\
\o1.5 & 0.556088 & 0.125156 &    0.689320 & 1.91987 \\
\o2.0 & 0.357380 & 0.116300 &    0.396190 & 1.56302 \\
\o3.0 & 0.210084 & 0.122228 &    0.150245 & 1.37844 \\
\o4.0 & 0.155000 & 0.140752 &    0.060109 & 1.39558 \\
\o5.0 & 0.127248 & 0.164516 &    0.019349 & 1.46241 \\
\o7.5 & 0.095384 & 0.232392 & $-$0.023752\m & 1.65805 \\
 10.0 & 0.081252 & 0.303784 & $-$0.044237\m & 1.83867 \\
 12.5 & 0.073272 & 0.375504 & $-$0.057057\m & 2.00116 \\
 15.0 & 0.068160 & 0.446580 & $-$0.065852\m & 2.14949 \\
 17.5 & 0.064608 & 0.516616 & $-$0.073915\m & 2.28498 \\
 20.0 & 0.062000 & 0.585444 & $-$0.077076\m & 2.41481 \\ \hline
\end{tabular}
\caption{\label{table:results}
Individual contributions plus the combined value
for the non-logarithmic constant $C_{\rm tot}(k/T)$ appearing in
the leading-order form (\ref{eq:Ctot}) for the
current-current correlator
(for lightlike momenta).
}
\end{table}

\begin{figure}[ht]
\hspace*{-20pt}
\hfill
\putbox{0.50}{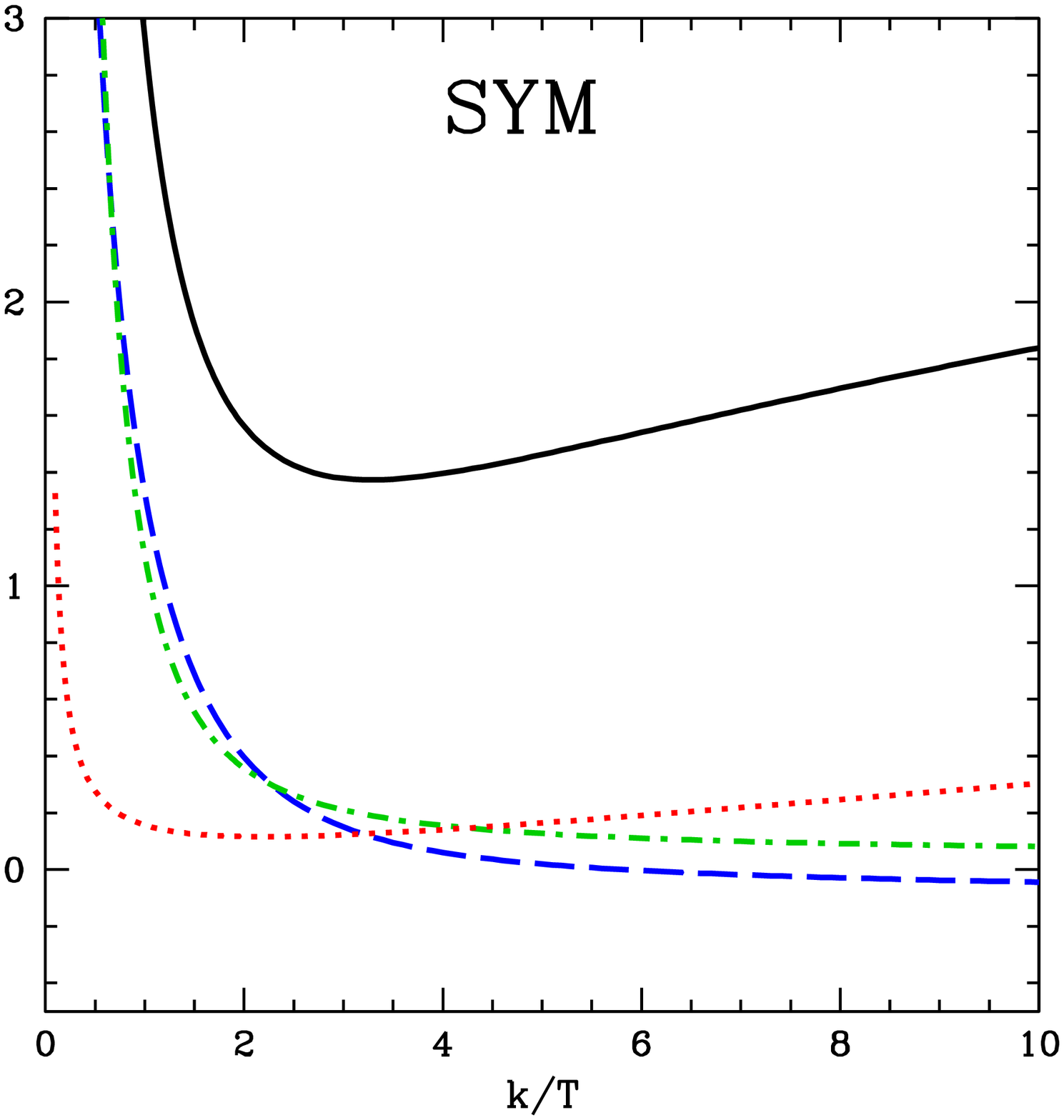} \hfill
\putbox{0.50}{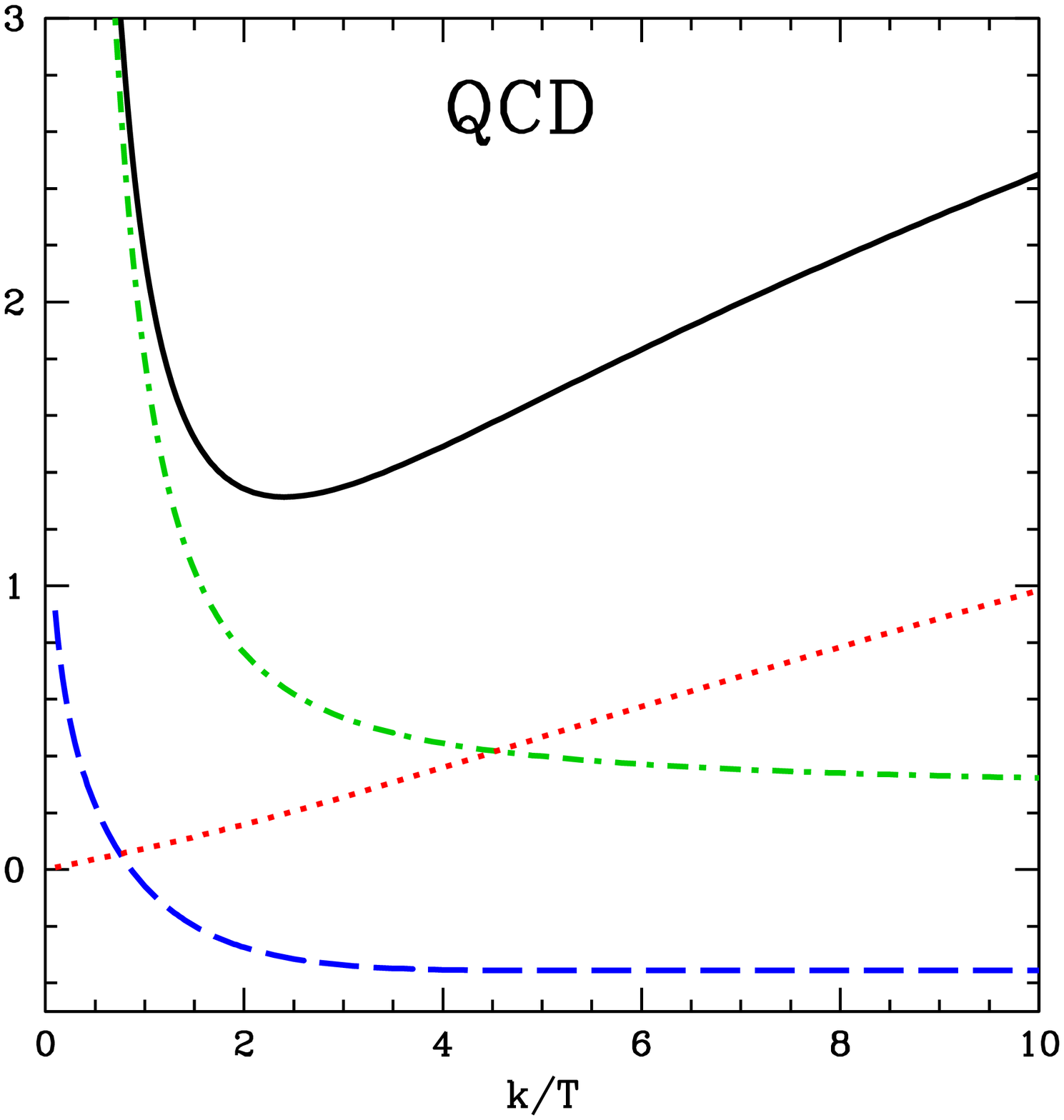} \hfill
\hspace*{-20pt}
\caption{\label{fig:results}
Photon emission contributions $C_{2\leftrightarrow 2}$ (blue dashed line),
$C_{\rm brem}$ (green dot-dashed), $C_{\rm pair}$ (red dotted),
and $C_{\rm tot}$ (solid black) as a function of $k/T$.
On the left are the results for $\Nfour$ SYM,
and on the right the corresponding contributions for three-flavor massless QCD.
}
\end{figure}

The various photon emission contributions are compared in
Figure \ref{fig:results}, for both $\Nfour$ SYM and for three-flavor QCD.
Notable differences between our SYM results and
the corresponding results for QCD \cite{AMY3} include the following.
\begin{itemize}
\item
The function $C_{2\leftrightarrow 2}(k/T)$ grows like $T/k$ for
frequencies small compared to $T$,
whereas in ordinary QCD the corresponding growth is only logarithmic in $T/k$.
This difference arises from Bose enhancement of scalar annihilation into a
photon and a gluon, a process not available in ordinary QCD.
Similarly, $C_{\rm pair}(k/T)$ rises at very small $k/T$ in SYM, but rapidly
goes to zero in QCD.  This reflects pair annihilation of scalars,
which is doubly Bose enhanced.
\item
At momenta of a few times $T$, 
inelastic processes are comparable in size to the
$2\leftrightarrow 2$ processes in QCD,
but are relatively less important in $\Nfour$ SYM.
This is because the $2\leftrightarrow 2$ processes arise mostly from
Compton-type scattering, which has a rate proportional to
the fermionic thermal mass, while inelastic processes arise because of
Coulomb scattering, with a rate proportional to the gauge boson thermal
mass.  In SYM the fermionic and gauge boson (asymptotic) thermal masses
are in 1:1 ratio, while in 3-flavor QCD they are in 4:9 ratio.
In addition, inelastic processes are suppressed in SYM by the larger
thermal mass appearing in the first term of \Eq{integral_eq}.  They
receive an extra contribution due to bremsstrahlung from scalars, but
this is subdominant for large momentum photons because the DGLAP kernel
coupling photons to scalars in \Eq{eq:finalint} is less efficient at
producing large momentum photons than the fermionic DGLAP kernel.
\item
At momenta of order $T$ or less, bremsstrahlung processes completely dominate
the emission rate in QCD, while in $\Nfour$ SYM the Bose-enhanced
scalar $\twotwo$ processes make a significant contribution down
to much smaller momenta.
\end{itemize}
Despite these difference, perhaps the most important feature is how
similar the result for $C_{\rm tot}(k/T)$ is between the two theories.
As seen in Fig.~\ref{fig:results}, the minimum value of $C_{\rm tot}$
is quite similar in the two theories.
The growth of $C_{\rm tot}(k/T)$ with increasing $k$ is a bit slower
in SYM, as compared with $\nf=3$ QCD, but in both theories
the asymptotic behavior%
\footnote
    {
    This asymptotic growth is slower than linear because of the
    effect of multiple soft scattering (or LPM suppression)
    limiting the formation time of the radiated photon.
    However, quite large values of $k/T$ are required to
    see this asymptotic behavior.
    }
is proportional to $(k/T)^{1/2}$.

At small frequency, for both QCD and SYM, the bremsstrahlung contribution
$C_{\rm brem}(k/T)$ behaves like $(T/k)^{3/2}$ (up to a log) and becomes
very large.
For any fixed photon frequency, this is the correct leading
weak-coupling behavior.
However, the limits of small coupling and small frequency do not commute.
For any given non-zero gauge coupling, this $(T/k)^{3/2}$
behavior cannot be valid all the way down to $k = 0$ because the
zero frequency limit of the correlator $\eta^{\mu\nu} C^<_{\mu\nu}(K)$
is proportional to the electrical conductivity
[as shown by Eq.~(\ref{eq:Kubo2})],
and this must be finite.
Our treatment of bremsstrahlung and pair annihilation requires
that photons emitted in response to a soft scattering event
be nearly collinear with the emitting charged particle.
This is valid for sufficiently weak coupling
at any given photon frequency,
but can fail for parametrically small frequency.
As discussed in some detail in Ref.~\cite{AMY2},
the relevant scale at which our analysis breaks down is
$k \sim \lambda^2 T$.
Below this scale, the growth of $C_{\rm brem}(k/T)$ must be
cut off and the correlator
$
	\eta^{\mu\nu}C^<_{\mu\nu}(K)
$
must approach a finite limiting value.

\subsection {Electrical conductivity}

The detailed behavior of the current-current correlator
for $k \sim \lambda^2 T$ is hard to compute.%
\footnote{%
    For momenta $\lambda^2 T \ll k \ll T$ the photon emission rate is
    determined by Eqs.~(\ref{eq:finalint}) and (\ref{integral_eq}),
    which (after using rotation invariance in the transverse plane) require
    the solution of a 1-dimensional integral equation in $p_\perp$.
    The complication at $k \sim \lambda^2 T$ is that the relevant values
    of $\p_\perp$ in \Eq{integral_eq} become $O(T)$, so the
    approximation $\p^2_\perp \ll p^2$ can no longer be made.  The
    problem then requires the solution of a 2-dimensional integral equation.
    For $k\ll \lambda^2 T$ the angular dependence becomes trivial and the
    problem is again reducible to
    a 1-dimensional integral equation.
    }
The
analogous calculation in ordinary QCD has not (yet) been performed
(though there are recent results for vanishing $\k$ but nonzero $k^0$
\cite{MooreRobert}).
In this regime, the formation time of the photon is so long that
the emitting particle should be thought of as undergoing diffusive motion,
not quasi-ballistic relativistic motion, during the emission event.
We will not analyze this regime here.
However, we can determine the value of the electrical conductivity,
to leading logarithmic accuracy,
and hence [via Eq.~(\ref{eq:Kubo2})]
the limiting $k\to 0$ value of the Wightman correlator.

As already discussed, the electrical conductivity is set by the
diffusion coefficient of charges.  The diffusion length is in turn
inversely related to the rate at which scatterings degrade a net
current.  The complication is
that it is a {\em functional} inverse, requiring the inversion of a
collision operator, which can be done approximately using variational
techniques.  We leave the detailed discussion to the literature
\cite{AMY1}; here we summarize the ideas and outline the differences
with respect to QCD.

Two types of scattering process are especially efficient at
scattering current-carriers, making a logarithmically enhanced
contribution to the collision operator.  The first comprises processes
with  $t$-channel gluon exchange (``Coulombic''
processes); these have an $s^2/t^2$ soft-divergent cross-section, cut off
by plasma effects; but since the initial and final state particles carry
the same charge, the effective scattering rate is only log divergent
({\em i.e.}, logarithmically sensitive to $T/\mD$).
The second comprises
processes involving a $t$-channel fermion exchange
(``Compton-like'' processes), which have an $s/t$ soft-divergent
cross-section and which completely re-orient the direction of the charge
carrier if the exchanged fermion is charged.  The rate of Coulombic
scattering can be determined directly from the presentation of
Ref.~\cite{AMY1}, though a new complication is that one must treat
separately the departure from equilibrium for scalars and fermions
(a complication already dealt with in Ref.~\cite{AMY1} in the context of shear
viscosity).  The
Compton-like cross-section affects both scalar and fermionic particles;
the cross-section is 16 times the one found by naively applying formulae
in Ref.~\cite{AMY1}, since each vertex can involve a gauge boson or one
of three scalar fields; however, only half of these scattering processes
destroy electrical current; the other half, in which a neutral fermion
is exchanged, flip the charge carrier between a scalar and a fermion.

Applying the technique presented in Ref.~\cite{AMY1}, taking into
account the differences just described,
one finds
\begin{equation}
\sigma = 1.28349 \, \frac{e^2 (\ncs{-}1) T}{\lambda^2  \,
  [\ln(\lambda^{-1/2}) + O(1)]} \, .
\label{sigma_weak}
\end{equation}
We have not evaluated the $O(1)$ constant.
Note that the $1/\lambda^2$ scaling (up to a log) is exactly
what one would find by simply cutting off the $k^{-3/2}$
small frequency growth of $C_{\rm brem}(k/T)$ at $k \sim \lambda^2 T$,
and inserting this into Eq.~(\ref{eq:Ctot}).
The behavior of $\eta^{\mu\nu} C_{\mu\nu}^<(K)$ near $k=0$ should
smoothly interpolate between the intercept of $4T\sigma/e^2$
and the form (\ref{eq:Ctot}) which is valid for $k \gg \lambda^2 T$;
both the limiting intercept, and Eq.~(\ref{eq:Ctot}), should
provide upper bounds on the actual value of the photon production rate.

\section{Discussion}
\label{sec:discussion}

Converting the differential photon emission rate
(\ref{eq:photon-rate-general}) into the emission rate
(per unit volume)
as a function of photon energy gives
\begin{equation}
    \frac{d\Gamma_\gamma}{dk}
    =
    \frac {\al}{\pi} \> k \> \eta^{\mu\nu} C^<_{\mu\nu}(K) \,.
\label{eq:dGam/dk}
\end{equation}
At low frequencies, the Wightman function $\eta^{\mu\nu} C^<_{\mu\nu}(K)$
approaches a constant proportional to the conductivity, as shown by the
Kubo formula (\ref{eq:Kubo2}), and hence $d\Gamma_\gamma/dk$
is linear in $k$ for small $k$,%
\footnote{
    In strongly coupled $\Nfour$ SYM theory,
    small $k$ means $k{\ll}T$.
    In weakly coupled $\Nfour$ SYM theory,
    small $k$ means $k{\ll}\lambda^2 T$,
    which is the inverse mean-free path for large-angle scattering.
    One should also keep in mind that even though
    $\eta^{\mu\nu} C_{\mu\nu}^< (k)$
    can in principle be computed for arbitrarily small $k$,
    the right-hand side of Eq.~(\ref{eq:dGam/dk})
    ceases to have the interpretation of the photon production rate if
    $k/T$ becomes comparable to either $e \nc$, or
    $(e^2 \ncs / \lambda)^{2/3}$.
    The first constraint reflects the fact that, due to electromagnetic
    corrections to the photon dispersion relation, photons no longer
    propagate through the plasma like nearly lightlike excitations
    if $k \lesssim e \nc T$.
    The second constraint reflects the scale where electromagnetic
    photon dispersion corrections
    can no longer be neglected in the integral equation (\ref{integral_eq}).
    See Ref.~\cite{AMY2} for details.
}
\begin{equation}
    \frac{d\Gamma_\gamma}{dk}
    =
    \frac{\sigma \, T}{\pi^2} \> k \,.
    \qquad \mbox{[small frequency]\kern-40pt}
\label{eq:smallfreq}
\end{equation}
At high frequencies, the Wightman function is Boltzmann suppressed,
as shown by the relation (\ref{KMS}) to the spectral density.
Therefore, in any equilibrium plasma,
the emission rate as a function of photon energy
must rise linearly from zero,
reach a maximal value, and eventually fall exponentially.

In weakly-coupled $\Nfour$ SYM theory,
the hydrodynamic regime in which (\ref {eq:smallfreq}) applies
is limited to $k \lsim \lambda^2 T$.
The slope $\sigma T/\pi^2$ is parametrically large,
as shown by the result (\ref{sigma_weak}) for the conductivity,
and the maximal value of $d\Gamma_\gamma/dk$
(which we have not evaluated quantitatively),
will be of order $\alphaEM (\ncs{-}1) T^3$ (up to a log of $\lambda$).
For photon momenta large compared to $\lambda^2 T$,
the analysis of section \ref{sec:weak-coupling} applies
and the photon emission spectrum may be expressed as
\begin{equation}
    \frac{d\Gamma_\gamma}{dk}
    =
    {\cal A} \> \frac {\alphaEM}{\pi^2} \> k \, \nfermi(k) \, m_\infty^2
    \big[
	\ln (T/m_\infty) + C_{\rm tot}(k/T)
    \big] \,,
\qquad \mbox{[large frequency]\kern-40pt}
\label{eq:dGam/dkB}
\end{equation}
with the coefficient $\mathcal A = \coeff 14(\ncs-1)$
for $\Nfour$ SYM (with our chosen charge assignments of $\pm\half$).
Here as before, $m_\infty^2=\lambda T^2$ describes the thermal
correction to hard fermion propagation in the medium.
Note that if one ignores the $k\gg \lambda^2 T$ condition on the
domain of validity of Eq.~(\ref{eq:dGam/dkB}) and uses this result
all the way down to $k = 0$,
then the $(k/T)^{-3/2}$ behavior of $C_{\rm brem}(k/T)$
will cause $d\Gamma_\gamma/dk$ to be singular at $k=0$, but because
of the explicit factor of $k$ in the formula (\ref{eq:dGam/dkB}) the
singularity is integrable
(and the energy-weighted spectrum is completely finite.)

In strongly coupled $\Nfour$ SYM theory, the photon emission spectrum
is obtained by inserting the $\lambda{=}\infty$ spectral density
(\ref{eq:trace-photons}) into Eqs.~(\ref{KMS}) and (\ref{eq:dGam/dk}), giving
\begin{equation}
  \frac{d\Gamma_\gamma}{dk} = \frac{\al\Nc^2 T^3}{16\pi^2}
  \frac{(k/T)^2}{e^{k/T}{-}1}
  \left|
  \ofo\!\Big( 1-\frac{(1{+}i)k}{4\pi T},1+\frac{(1{-}i)k}{4\pi T};
       1{-}\frac{ik}{2\pi T};-1 \Big)
  \right|^{-2} .
\label{eq:sc-sym-rate}
\end{equation}
This is an exact expression (in the large $\nc$, large $\lambda$ limit),
valid for for all photon energies,
both large and small.
Equation (\ref{eq:sc-sym-rate}) naturally reproduces the
small-momentum form (\ref{eq:smallfreq}) in the hydrodynamic limit,
with the electrical conductivity given by (\ref{eq:sigma}).
Thus the slope of $d\Gamma_\gamma/dk$ at small momentum
is coupling-independent in the limit of large coupling,
and is parametrically smaller than the corresponding slope
in the weakly coupled theory.
The maximum of $d\Gamma_\gamma/dk$ is attained at
$k_{\rm max}\approx1.48479\,T$,
with the maximal rate of
\begin{equation}
  \left(\frac{d\Gamma_\gamma}{dk}\right)_{\rm max} \approx\,
  0.01567\, \al \Nc^2 T^3 \,.
\end{equation}
At arbitrary values of the coupling, one must have
$(d\Gamma_\gamma/dk)_{\rm max} = f(\lambda)\, \al \Nc^2 T^3$,
where $f(\lambda)$ interpolates between the strong-coupling
result $f(\lambda{\to}\infty)\approx0.01567$, and the weak-coupling
maximal intensity.
To date, the weak-coupling expression for $f(\lambda)$
has not been calculated, in any gauge theory.
At large momenta, $k{\gg}T$, the photon rate in
strongly coupled SYM theory decays as $k^{5/3} e^{-k/T}$, a stronger
power than the $k^{3/2} e^{-k/T}$ rate one finds in the extreme large $k$
limit of the weak-coupling calculation (in which 
$C_{\rm pair}(k/T) \sim k^{1/2}$ due to LPM suppression).

\begin{figure}[ht]
  \vspace*{-10pt}
  \includegraphics[width=3.8in]{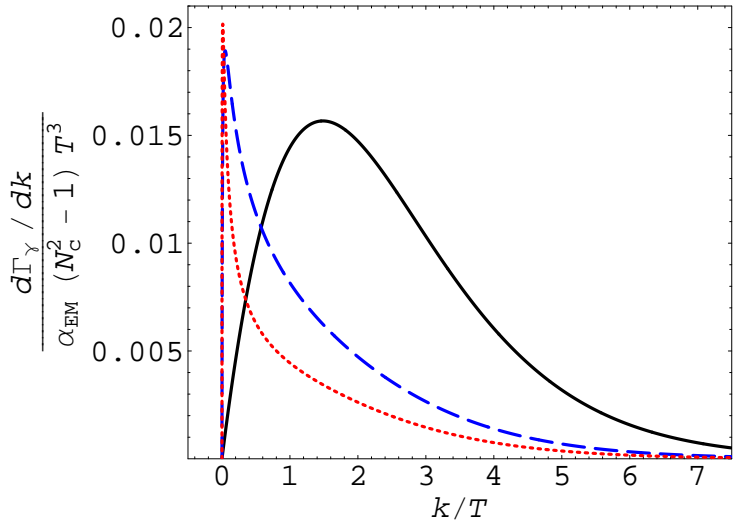}
  \vspace*{-20pt}
\caption{
  Photo-emission spectrum $d\Gamma_\gamma/dk$,
  divided by $\alphaEM (\ncs-1) T^3$,
  in $\Nfour$ supersymmetric
  Yang-Mills theory
  for
  $\lambda=\infty$ (solid black curve),
  $\lambda =0.5$ (dashed blue), and
  $\lambda = 0.2$ (dotted red).
  As explained in footnote \ref{fn:interp},
  the weak-coupling curves interpolate between the rising
  small-frequency result (\ref{eq:smallfreq}),
  valid for $k \lesssim \lambda^2 T$,
  and the falling large-frequency result (\ref{eq:dGam/dkB}), valid for
  $k \gg \lambda^2 T$;
  the precise height of the sharp narrow peak is not known.
}
\label{fig:comparison1}
\end{figure}

Fig.~\ref{fig:comparison1} illustrates how
the photo-emission spectrum evolves as $\lambda$ increases.%
\footnote
    {
    The weak-coupling curves in Figs.~\ref{fig:comparison1}
    were generated using a smooth interpolation between
    the small frequency form (\ref{eq:smallfreq}) and
    the form (\ref{eq:dGam/dkB}) for $O(1)$ values of $k/T$,
    with the unknown $O(1)$ constant in the conductivity
    (\ref{sigma_weak}) set to $\half \ln \coeff 92$.
    Ref.~\cite{AMY4}, which evaluated the complete leading-order
    flavor diffusion constant (or equivalently, the conductivity)
    in various QED and QCD-like theories, found that the correct
    constant to be added to the log equals this value to within $\pm 8\%$
    for a variety of non-Abelian theories with different matter content.
    So this is our best guess for the appropriate value for SYM theory.
    In addition, the photon rate in strongly coupled SYM theory
    was plotted with $\Nc^2$ replaced with $\Nc^2{-}1$ in the
    expression (\ref{eq:sc-sym-rate}).
    \label{fn:interp}
    }
The slope at $k=0$ (proportional to the conductivity) decreases,
the position and width of the hydrodynamic peak
(both proportional to $\lambda^2 T$) increase,
and the amplitude of the spectrum for $k/T \gsim 1$ increases
[due to the factor of $m_\infty^2 \propto \lambda$ in Eq.~(\ref {eq:dGam/dkB})].
Figure~\ref{fig:comparison1} also shows the strong coupling result
(\ref{eq:sc-sym-rate}).
The strong coupling curve differs primarily in having the
temperature $T$, and not some smaller scale,
set the width of the hydrodynamic regime
(in which $d\Gamma_\gamma/dk$ is approximately linear in $k$).
As a result, there is a broad maximum in the strong coupling
spectrum at $k_{\rm max}\approx 1.5 T$.
At sufficiently small frequencies, the photon production rate is largest
in the most weakly coupled theory.
This is because, at these frequencies, the photon wavelength is larger
than the free path of the particles involved, so the charges are
effectively diffusing; weak coupling means faster diffusion and
therefore more current on such long scales.
The cross-over point, below which the weak-coupling rate exceeds the
($\lambda$ independent) strong-coupling rate, scales as $\lambda^{2/3} T$.
At large frequency, the production rate is greatest in the
strongly coupled theory. This is because the spectrum, for weak coupling,
is proportional to the gauge coupling $\lambda\ll 1$
[appearing in Eq.~(\ref{eq:dGam/dkB}) as the factor $m_\infty^2 =
\lambda T^2$],
while the spectrum in the strong coupling limit has no such suppression.
The results are clearly consistent with an expectation of smooth
evolution between the weak and strong coupling regimes.%
\footnote
    {
    The large momentum behavior of the spectral density
    $\eta^{\mu\nu} \chi_{\mu\nu}(K)$, for lightlike momenta,
    differs quantitatively between weak and strong coupling,
    growing proportional to $k^{1/2}$
    for weak coupling (due to $C_{\rm pair}(k/T)$),
    and proportional to $k^{2/3}$ for strong
    coupling, as seen in Eq.~(\ref{f_grav}).
    However, one can easily imagine that the true asymptotic behavior
    is a coupling-dependent power-law $k^{\nu(\lambda)}$
    with an exponent $\nu(\lambda)$
    which smoothly interpolates between the two limiting values.
    }

It is instructive to compare our weak-coupling result
for $d\Gamma_\gamma/dk$ with the corresponding result for QCD \cite{AMY3}.
But before this can be done in a meaningful fashion,
we must first address the question of what normalization of the
$U(1)$ current in SYM-EM will best mimic the electromagnetic physics
of a real QGP.
One can consider various different criteria
for fixing the charge (or current) normalization
in order to compare results between different theories.%
\footnote
    {
    For example,
    requiring equality of the sum of squares of
    electric charges of all charged fields might seem natural.
    Or one could require equality of the EM charge susceptibility,
    which measures mean square fluctuations in electric charge density.
    These criteria are different (and both differ from our chosen
    condition).
    }
But for our purposes
(involving the comparison of electromagnetic emission phenomena),
the most natural criterion for fixing the normalization of the
EM current in SYM-EM is to require
that the dilepton emission spectra agree at large invariant mass.
This amounts to demanding that the leading behavior of the current-current
spectral density at large time-like momentum coincide in the two theories.
This provides a simple criterion which will fix the normalization
of the current, independent of the interaction strength in SYM,
because all medium dependent corrections to the spectral density
$\chi_\mu^\mu(K)$ vanish exponentially when $|K^2| \gg T^2$.
Coinciding behavior of current-current spectral density
at large timelike momenta also implies coinciding behavior
at large spacelike momenta, and hence this criterion is the same
as demanding that the leading short distance behavior of the
$\langle J_\mu^{\rm EM}(x) J_\nu^{\rm EM}(0) \rangle$ correlator
agree between theories.
(Hence this criterion is the same as that used in Ref.~\cite{wall3}.)

The particular choice of current we made in Eq.~(\ref{eq:Sint})
happens to make the fermionic contribution to the large momentum
behavior of the current-current spectral density the same as in
QCD with three flavors (and $\Nc=3$).
This merely reflects the fact that the sum of squares of the fermion
charges coincide.
But the scalars of $\Nfour$ SYM contribute
half as much as the fermions
to the high momentum spectral density.
Consequently, to satisfy the condition of coinciding large
momentum behavior of the spectral density (for $\Nf = \Nc = 3$),
the $U(1)$ current used in our SYM calculations should be rescaled
by a factor of $\sqrt{2/3}$.
This we do in the following comparisons.

\begin{figure}
  \begin{picture}(0,0)(0,0)
  \put(160,160){$\alphas^{\rm QCD} = \alphas^{\rm SYM}$}
  \end{picture}
  \includegraphics[width=3.3in]{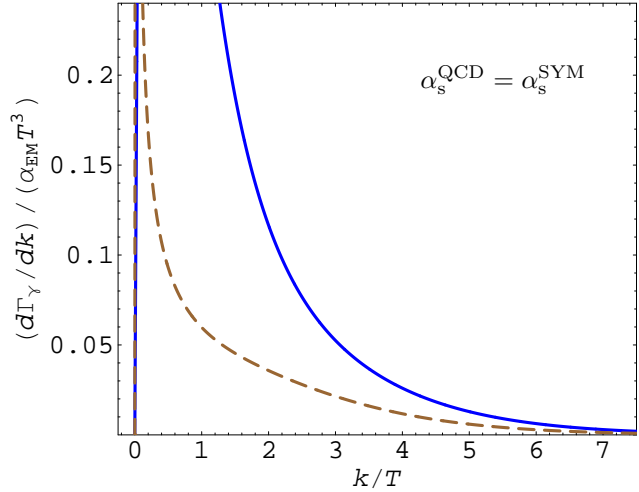}
  \vspace*{-10pt}
\caption{
  Comparison of leading-order photo-emission spectra
  for three-flavor massless QCD and $\Nfour$ SYM
  (for $\nc = 3$), when the gauge coupling
  has the same value, $\alphas= 0.1$, in both theories.
  The upper solid curve shows the SYM result,
  while the lower dashed curve is the QCD result.
}
\label{fig:comparison2}
\end{figure}

Once the normalization of the $U(1)$ current is fixed,
the only other adjustable parameter in the SYM emission spectra
is the value of the gauge (or 't Hooft) coupling $\lambda$.
If one compares photon emission in three-flavor QCD and $\Nfour$ SYM,
at $\Nc = 3$ and the same value of the gauge coupling in both theories,
then the SYM photo-emission rate is dramatically larger than
the QCD rate, as shown in Fig.~\ref{fig:comparison2}.
This difference arises because the scattering rate in the SYM
plasma is much higher than in QCD, owing to the larger number of
matter fields and the fact that they are in the adjoint representation.
In particular,
the rate at which a quark undergoes photon producing Compton-type
scattering is proportional to $m_\infty^2$, which is 9 times larger in
SYM than in QCD (for equal values of the gauge coupling).
The rate of Coulomb scattering, important in
bremsstrahlung, is proportional to $m_{\rm D}^2$, which is 4 times
larger.

\begin{figure}
  \hspace*{-30pt}
  \begin{picture}(0,0)(0,0)
  \put(150,160){$\mD^{\rm QCD} = \mD^{\rm SYM}$}
  \end{picture}
  \includegraphics[width=3.2in]{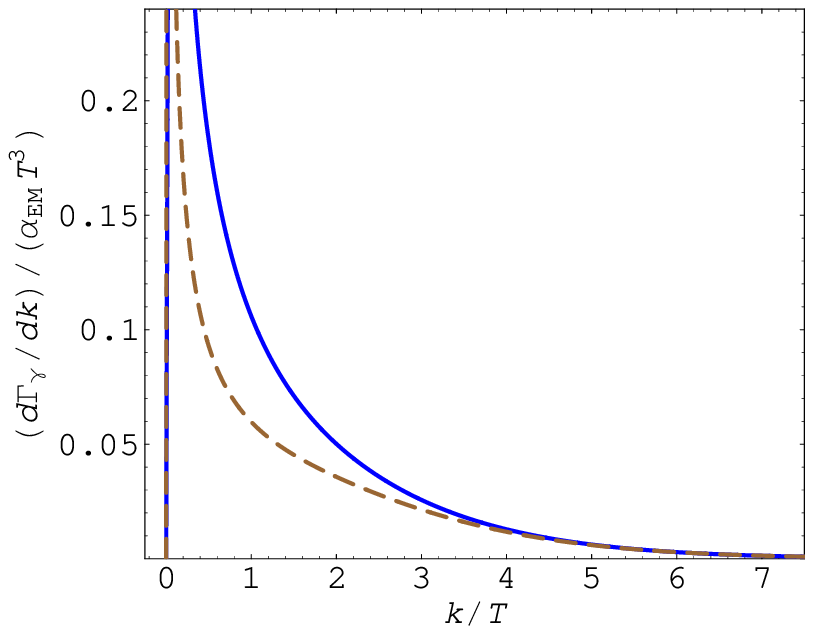}
  \hspace*{ 5pt}
  \begin{picture}(0,0)(0,0)
  \put(150,160){$m_\infty^{\rm QCD} = m_\infty^{\rm SYM}$}
  \end{picture}
  \raisebox{0pt}{\includegraphics[width=3.2in]{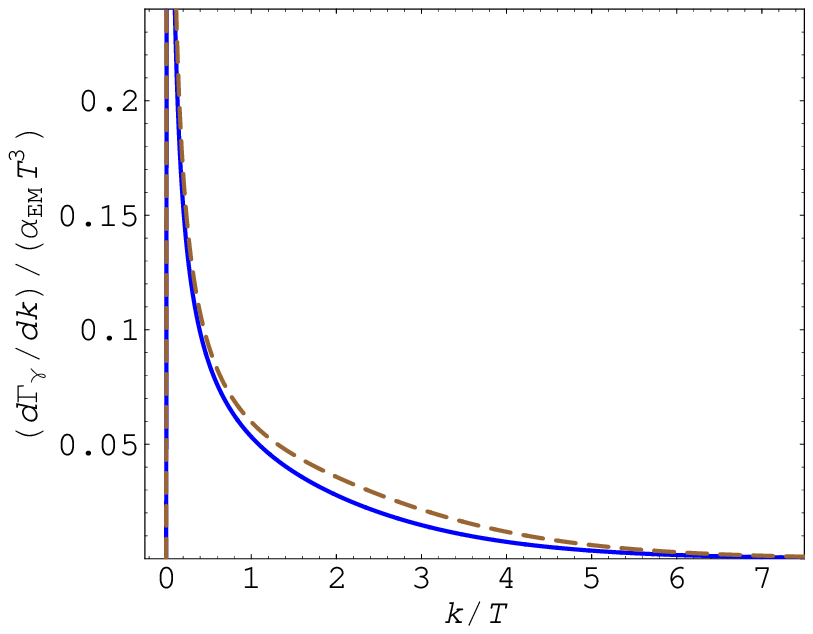}}
  \hspace*{-30pt}
  \vspace*{-10pt}
\caption{
  Comparison of leading-order photo-emission spectra
  for three-flavor massless QCD (dashed curve) and $\Nfour$ SYM (solid curve)
  at $\nc = 3$.
  Left: equal values of the Debye mass $\mD$ in both theories,
  corresponding to
  $\alphaSYM = 0.025$ in SYM
  and $\alphas = 0.1$ in QCD.
  Right: equal values of the asymptotic fermion mass $m_\infty$,
  corresponding to
  $\alphaSYM = 0.011$ in SYM and $\alphas = 0.1$ in QCD.
}
\label{fig:comparison3}
\end{figure}

Because of this difference in scattering rates,
it is more appropriate to compare the two
theories with the SYM gauge coupling adjusted to give
either the same value of the Debye mass $m_{\rm D}$
(in which case $\lambda^{\rm SYM} = \coeff 14 \lambda^{\rm QCD}$),
or the same value of $m_\infty$ for the fermions as in QCD
(in which case $\lambda^{\rm SYM} = \coeff 19 \lambda^{\rm QCD}$).
Both of these comparisons are shown in Fig.~\ref{fig:comparison3}.
With coinciding values of the Debye mass,
shown on the left of Fig.~\ref{fig:comparison3},
the two spectra
are nearly identical at high momenta, $k \gtrsim 3T$,
while the SYM rate is larger at lower momenta
(about 75\% larger at $k/T = 1$).
When the two theories are compared at coinciding values of the
asymptotic fermion mass $m_\infty$, 
as shown on the right hand plot of Fig.~\ref{fig:comparison3},
the resulting curves are remarkably similar,
with the SYM spectrum just a bit below the QCD result.

A major motivation of this work was the hope that one would be able
to translate knowledge of the photon production rate in strongly
coupled SYM into a useful prediction for the rate in strongly coupled QCD.
This seems reasonably plausible given that
(for comparable numbers of charged quarks)
the weak coupling photon spectra of the two theories
are quite similar,
as shown in Fig.~\ref{fig:comparison3} ---
provided one scales the 't Hooft coupling of SYM
relative to that of QCD by a factor somewhere in the range 4--9.
However, it is not clear if the quark-gluon plasma produced in
heavy ion collisions can be well modeled by SYM plasma in the
asymptotically strongly coupled $\lambda \to \infty$ limit.
A comparison of \Eq{eq:trace-photons} and \Eq{eq:Ctot}
at, say, $k=5T$, shows that the weak coupling result approaches
within 20\% of the strong coupling result at $\lambda^{\rm SYM} \approx 4$.
Taking this as a guess for the beginning of the
region where the asymptotic strong coupling result
provides a decent approximation,
and applying the above rescaling between SYM and QCD gauge couplings
yields $\alphas^{\rm QCD}$ in the range 0.4--1.0,
somewhat larger than the values of $\alphas \approx 0.3$--0.5
commonly thought to be relevant in
the QGP formed in heavy ion collisions.
Therefore, it may be best to view the photon production rate
in infinitely strongly-coupled SYM,
for $k/T > 1$,
as an upper bound on what we expect the
photon production rate from real QGP to be.
Nevertheless, it will be interesting to see if incorporation of the
strong coupling SYM
spectral functions into models of photon production in heavy ion
collisions improves the comparison with data.
We have recently learned that efforts to do so are underway
\cite{Jan-e}.

\begin{figure}
  \hspace*{-35pt}
\psfrag{sqrtK2}{\raisebox{-1ex}
  {\footnotesize\hspace{-0.2cm}$\sqrt{\wn^2-\qn^2}$}}
\psfrag{SYM1}{\raisebox{1ex}{\hspace{-0.4cm}SYM, $\lambda{=}0$}}
\psfrag{SYM2}{\raisebox{1ex}{\hspace{-0.4cm}SYM, $\lambda{=}\infty$}}
\psfrag{delChimmrel}{$\Delta\chimm/\chimm(T{=}0)$}
  \includegraphics[width=3.5in]{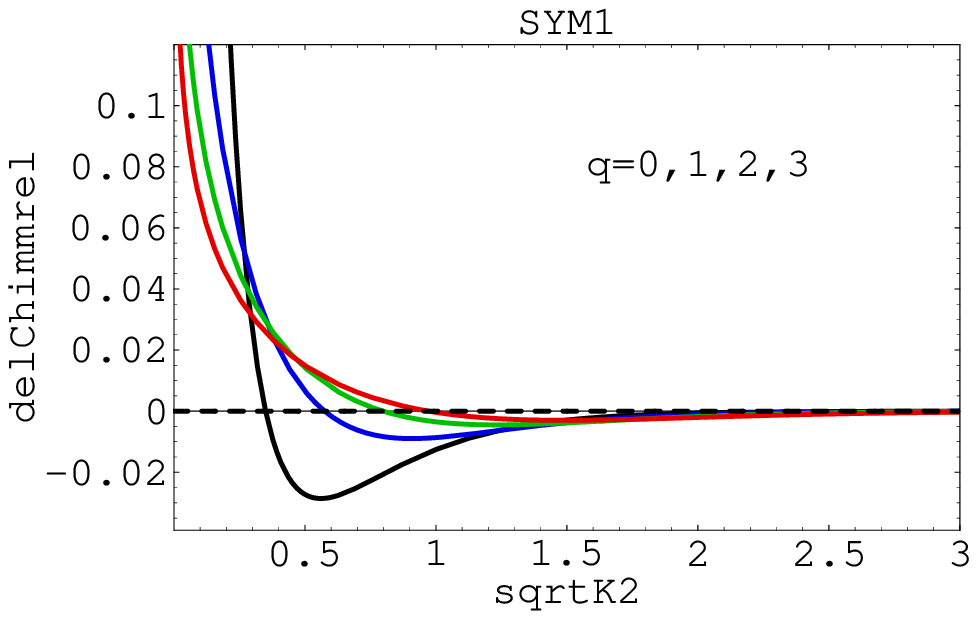}
  \hspace*{-15pt}
  \includegraphics[width=3.5in]{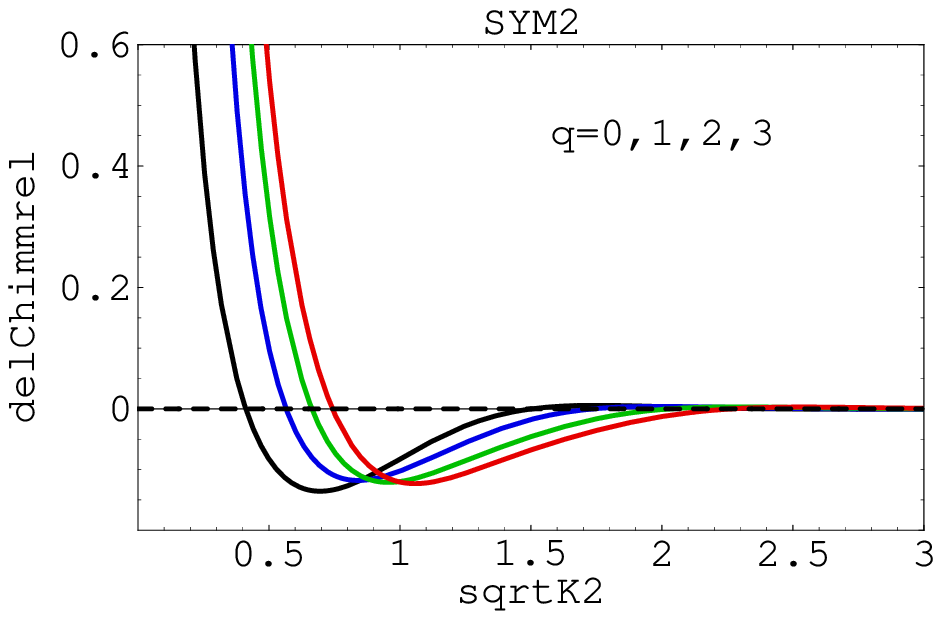}
  \hspace*{-30pt}
\caption{
  Relative size of the thermal corrections to the trace of the
  spectral function,
  $\Delta\chimm(k^0,k)$ divided by $\chimm(k^0,k)|_{T=0}$,
  as a function of $|K|/(2\pi T) = \sqrt{w^2-q^2}$
  for $q \equiv k/(2\pi T)$ = 0 (black), 1 (blue), 2 (green), and 3 (red).
  The left panel shows the weak-coupling $\lambda\to 0$ result,
  while the right panel shows the strong coupling $\lambda\to\infty$ result.
}
\label{fig:dilepton}
\end{figure}

Regarding dilepton emission, our results have a simple and
more positive implication.
Sufficiently deep in the timelike region,
thermal corrections to the spectral function become very small for
both weak and strong coupling.
This is shown explicitly in Fig.~\ref{fig:dilepton},
which plots the relative correction to the zero temperature
spectral density as a function of $\sqrt{-K^2}$. 
If $|K^2| \ge (2\pi T)^2$,
then thermal corrections to spectral function
at weak coupling are under 2\%.
At strong coupling the corrections are larger, but nevertheless
no more than 15\% in this regime.
Since the zero temperature spectral density is independent of coupling,
as discussed in Section~\ref{sec:strong-coupling}, this means
that the dilepton spectrum is nearly identical at weak and strong coupling,
as long as the invariant mass of the pair is above $2\pi T$.
This is also consistent with the modest size of the next-to-leading order
weak-coupling result in QCD \cite{Majumder},
which is a relative correction of
$-\coeff {8}{9}(\alphas/2\pi) (2\pi T)^2/K^2$.
Therefore, for large invariant
mass dilepton pairs, it is undoubtedly an excellent approximation to use
the lowest-order production rate calculation even when the coupling is
strong.

Our results for photon and dilepton emission rates
at strong coupling can be extended in a number of obvious ways.
One question is how the emission spectra at strong coupling change
if the field theory deviates from the conformally symmetric $\Nfour$ SYM case.
As discussed above, perturbative rates of photon production
in conformal $\Nfour$ SYM and in (non-conformal) QCD
do not differ significantly, when compared at the same value
of the thermal fermion mass.
This suggests that the presence or absence of conformal symmetry does
not play a decisive role in determining the perturbative photon spectrum.
AdS/CFT can provide a similar comparison at strong coupling.
For example, it should be possible to compute photon and dilepton
emission spectra in both mass-deformed $\Nfour$ SYM \cite{N=2*},
and in $\Nfour$ SYM at non-zero chemical potential,
at strong coupling.
A further question is related to the coupling constant
dependence of the emission rates.
It would be interesting to see how the
spectrum shown in Fig.~\ref{fig:comparison1}
evolves when $O(\lambda^{-3/2})$ corrections \cite {Buchel:2004di} are taken
into account.
It would also be interesting to extend our analysis
to theories with matter fields in the
fundamental representation of the gauge group.
Adding fundamental representation matter to
$\Nfour$ SYM corresponds, in the gravity dual,
to the addition of D7 branes embedded in
the $AdS_5 \times S^5$ geometry \cite{KarchKatz}.
In such theories,
it is natural to regard a flavor symmetry current
of the fundamental matter fields as the electromagnetic current.
Analyzing vector fluctuations on the $D7$ brane
would then allow one to compute, at strong coupling,
photon production from fundamental representation quarks
of arbitrary mass added to the $\Nfour$ SYM plasma.


\begin{acknowledgments}
Chris Herzog and Andreas Karch are thanked for helpful conversations.
The work of P.K.\ was supported in part by the
U.S.\ National Science Foundation under Grant No. PHY99-07949.
The work of L.Y.\ was supported in part by the
U.S.\ Department of Energy under Grant No.~DE-FG02-96-ER-40956.
The work of G.M.\ and of S.C.\ was supported by the
National Sciences and Engineering
Research Council of Canada, and by le Fonds Nature et Technologies du
Qu\'ebec.
Research at Perimeter Institute is supported in part
by the Government of Canada through NSERC and by the
Province of Ontario through MEDT.
\end{acknowledgments}

\appendix
\section{Asymptotics of the spectral function}
\label{app:photons-asympt}

For null momenta ($w{=}q$),
Eq.~(\ref{eq:Ex}) for the transverse component
of the electric field has the form
\begin{equation}
E_\perp '' - \frac{2 u}{f}\, E_\perp ' + \frac{\wn^2 u}{f^2}\, E_\perp =0\,.
\label{equation}
\end{equation}
The solution obeying the incoming wave boundary condition at the horizon ($u=1$) is 
\begin{equation}
E_\perp (u) = (1-u)^{-i \wn/2} (1+u)^{-\wn/2}  \,
\ofo \!\left( - \half{(1{+}i)\wn}\,, 1 {-} \half{(1{+}i)\wn};1 {-} i \wn 
;\half{(1-u)}\right)\,.
\label{exact}
\end{equation}
We are interested in asymptotics of the solution 
(\ref{exact}) for large and small values of $\wn$.
To the best of our knowledge, appropriate asymptotic expansions
of the hypergeometric function are unavailable in the literature.
However, such expansions can be readily
 derived from the differential equation (\ref{equation}) 
following the approach of Ref.~\cite{hep-th/0104065}.

For $\wn \gg 1$, we 
use the 
Langer-Olver method \cite{olver} for constructing uniform asymptotic 
expansions (a version of the WKB approximation). 
Introducing new variables,
\begin{equation}
E_\perp (u)  = \frac{1}{\sqrt{-f(u)}}\; y(u)\,, \qquad \qquad  x = -u\,,
\end{equation}
one can rewrite Eq.~(\ref{equation}) as 
\begin{equation}
y''(x) = \frac{\wn^2 x -1}{(1-x^2)^2}\;   y(x)\,.
\label{eqx}
\end{equation}
For $\wn\rightarrow \infty$, the dominant term on the right-hand side
of Eq.~(\ref{eqx})
has a simple zero at $x=0$ and thus according to Ref.~\cite{olver} 
the asymptotics can be expressed in terms of Airy functions. 
Moreover, since  the coefficients of Eq.~(\ref{eqx}) satisfy 
the conditions of Theorem 3.1 of Chapter XI
 in Ref.~\cite{olver}, one is guaranteed to
have a uniform asymptotic expansion for all $u\in [0,1]$.
The asymptotic expansion is
\begin{equation}
E_\perp(u) \sim \frac{C(\wn)}{\sqrt{-f(u)}} \; \hat{f}^{-1/4}(-u)\; \mbox{Ai}
\left(\wn^{2/3} \zeta (-u)\right) +\cdots\,,
\label{asex}
\end{equation}
where $\mbox{Ai}(z)$ is the Airy 
function,\footnote{The choice of the Airy function $\mbox{Ai}(z)$ rather
 than $\mbox{Bi}(z)$ is dictated by 
the incoming wave boundary conditions at the horizon.} 
\begin{equation}
\zeta (x) \equiv \frac{3^{2/3}}{2^{4/3}} \left( i \pi - 2 \arctan{\sqrt{x}}
+\log{ \frac{\sqrt{x}+1}{\sqrt{x}-1}}\right)\,, \qquad \hat{f}(x) \equiv \frac{x}{(1-x^2)^2 \, \zeta(x)}\,,
\end{equation}
and the ellipses denote corrections that can
be systematically computed \cite{olver}.
The normalization constant,
\begin{equation}
C(\wn)= 2 \sqrt{\pi}
 e^{i \pi/4} \, \wn^{1/6} \, 2^{-i \wn - \wn/2}\, e^{i \pi \wn/4}\,,
\end{equation}
 is chosen  
in such a way that the asymptotic expansion (\ref{asex}) 
coincides with the exact solution (\ref{exact}) as $u\rightarrow 1$,
where
 $E_\perp \rightarrow  
 (1{-}u)^{-i \wn/2} \, 2^{-\wn/2}$. 
Using the asymptotic solution (\ref{asex}), 
for the retarded correlators we find
\begin{equation}
\Pi^T(\wn{=}\qn) \sim - \frac{N_c^2 T^2}{8}\,
 \frac{(-\wn)^{2/3} \, 3^{1/3}\, \Gamma (2/3)}{\Gamma(1/3)}\,, \qquad \wn \gg 1\,.
\end{equation}
Correspondingly, for the trace of the spectral function we obtain
\begin{equation}
\chimm (\wn{=}\qn) \sim \frac{N_c^2 T^2 \, \wn^{2/3}}{4}\,
 \frac{ 3^{5/6} \, \Gamma (2/3)}{\Gamma (1/3)} \,, \qquad \wn \gg 1\,.
\end{equation}

In the low-frequency limit, one can solve Eq.~(\ref{equation}) perturbatively
 using $\wn \ll 1$ as a small parameter. Since this procedure
 is well known (see, for example, Refs.~\cite{hep-th/0104065,hep-th/0205052}),
 we omit the details.
The retarded correlator for $\wn \ll 1$ is given by
\begin{equation}
\Pi^T(\wn{=}\qn) = - \coeff i8 {N_c^2 T^2 \, \wn}
\left[ 1 +i \wn \log{2} -\coeff 1{12} {\pi^2} \wn^2
 + O(\wn^3)\right] ,
\end{equation}
and the resulting trace of the spectral function is
\begin{equation}
\chimm (\wn{=}\qn) = \half{N_c^2 T^2 \,
 \wn}\left[ 1 - \coeff 1{12}{\pi^2}\, \wn^2 + O(\wn^4)\right] \,.
\end{equation}


\begin{thebibliography}{99}

\bibitem{photons-rhic}
For a review, see
  P.~Stankus,
  {\it Direct photon production in relativistic heavy-ion collisions,}
  Ann.\ Rev.\ Nucl.\ Part.\ Sci.\  {\bf 55}, 517 (2005).

\bibitem{RHIC_photon1}
S.~S.~Adler {\it et al.}  [PHENIX Collaboration],
  {\it Measurement of identified $\pi_0$ and inclusive photon $v_2$
      and implication  to
  the direct photon production in $s(NN)^{1/2} = 200$ GeV Au+Au collisions,}
  Phys.\ Rev.\ Lett.\  {\bf 96}, 032302 (2006),
  {\tt nucl-ex/0508019};

\bibitem{RHIC_photon2}
S.~S.~Adler {\it et al.}  [PHENIX Collaboration],
  {\it Centrality dependence of direct photon production in $s(NN)^{1/2} =
  200$ GeV Au+Au collisions,}
  Phys.\ Rev.\ Lett.\  {\bf 94}, 232301 (2005),
  {\tt nucl-ex/0503003}.

\bibitem{Vogelsang}
L.~E.~Gordon and W.~Vogelsang,
  {\it Polarized and unpolarized prompt photon production beyond the leading
  order,}
  Phys.\ Rev.\ D {\bf 48}, 3136 (1993).

\bibitem{AMY3}
P.~Arnold, G.~D.~Moore and L.~G.~Yaffe,
  {\it Photon emission from quark gluon plasma:
  Complete leading order  results,}
  JHEP {\bf 0112}, 009 (2001),
  {\tt hep-ph/0111107}.

\bibitem{Le-Bellac}
  See, for example
  M.~Le~Bellac,
  {\sl Thermal Field Theory,}
  Cambridge, 1996.

\bibitem{rates-pert2}
  P.~Aurenche, F.~Gelis, G.~D.~Moore and H.~Zaraket,
  {\it Landau-Pomeranchuk-Migdal resummation for dilepton production,}
  JHEP {\bf 0212}, 006 (2002),
  {\tt hep-ph/0211036}.

\bibitem{rates-lattice}
  F.~Karsch, E.~Laermann, P.~Petreczky, S.~Stickan and I.~Wetzorke,
  {\it A lattice calculation of thermal dilepton rates,}
  Phys.\ Lett.\ B {\bf 530}, 147 (2002),
  {\tt hep-lat/0110208}.

\bibitem{Gupta}
  S.~Gupta,
  {\it The electrical conductivity and soft photon emissivity
  of the QCD plasma,}
  Phys.\ Lett.\ B {\bf 597}, 57 (2004),
  {\tt hep-lat/0301006}.

\bibitem{hep-lat/0607012}
  G.~Aarts, C.~Allton, J.~Foley, S.~Hands and S.~Kim,
  {\it Meson spectral functions at nonzero momentum in hot QCD,}
  {\tt hep-lat/0607012}.

\bibitem{rates-review}
  J.~P.~Blaizot and F.~Gelis,
  {\it Photon and dilepton production in the quark-gluon plasma:
  Perturbation theory vs lattice QCD,}
  Eur.\ Phys.\ J.\ C {\bf 43}, 375 (2005),
  {\tt hep-ph/0504144}.

\bibitem{AGMOO}
For a review, see
  O.~Aharony, S.~S.~Gubser, J.~M.~Maldacena, H.~Ooguri and Y.~Oz,
  {\it Large $N$ field theories, string theory and gravity,}
  Phys.\ Rept.\  {\bf 323}, 183 (2000),
  {\tt hep-th/9905111}.

\bibitem{spectral}
  P.~Kovtun and A.~Starinets,
  {\it Thermal spectral functions of strongly coupled $\Nfour$ supersymmetric
  Yang-Mills theory,}
  Phys.\ Rev.\ Lett.\  {\bf 96}, 131601 (2006),
  {\tt hep-th/0602059}.

\bibitem{derek}
  D.~Teaney,
  {\it Finite temperature spectral densities of momentum and
  $R$-charge correlators in $\Nfour$ Yang Mills theory,}
  Phys.\ Rev.\ D {\bf 74}, 045025 (2006)
  {\tt hep-ph/0602044}.

\bibitem{recipe}
  D.~T.~Son and A.~O.~Starinets,
  {\it Minkowski-space correlators in AdS/CFT correspondence: Recipe and
  applications,}
  JHEP {\bf 0209}, 042 (2002),
  {\tt hep-th/0205051}.

\bibitem{hep-th/0205052}
  G.~Policastro, D.~T.~Son and A.~O.~Starinets,
  {\it From AdS/CFT correspondence to hydrodynamics,}
  JHEP {\bf 0209}, 043 (2002),
  {\tt hep-th/0205052}.

\bibitem{NS}
  A.~Nunez and A.~O.~Starinets,
  {\it AdS/CFT correspondence, quasinormal modes, and thermal
  correlators in $\Nfour$ SYM,}
  Phys.\ Rev.\ D {\bf 67}, 124013 (2003),
  {\tt hep-th/0302026}.

\bibitem{quasipaper}
  P.~K.~Kovtun and A.~O.~Starinets,
  {\it Quasinormal modes and holography,}
  Phys.\ Rev.\ D {\bf 72}, 086009 (2005),
  {\tt hep-th/0506184}.

\bibitem{Anselmi:1997am}
   D.~Anselmi, D.~Z.~Freedman, M.~T.~Grisaru and A.~A.~Johansen,
  {\it Nonperturbative formulas for central functions of supersymmetric gauge
  theories,}
  Nucl.\ Phys.\ B {\bf 526}, 543 (1998),
  {\tt hep-th/9708042}.

\bibitem{yellow-book}
  See, for example,
  C.~M.~Bender, S.~A.~Orszag,
  {\it Advanced Mathematical Methods for Scientists and Engineers,}
  Springer, 1999.

\bibitem{abramowitz}
  See, for example, Eq. (15.3.4) of
  M.~Abramowitz and I.~A.~Stegun, Ed.,
  {\it Handbook of Mathematical Functions,}
  Dover, New York, 1970.

\bibitem{Jpsi-lattice1}
  T.~Umeda, K.~Nomura and H.~Matsufuru,
  {\it Charmonium at finite temperature in quenched lattice QCD,}
  Eur.\ Phys.\ J.\ C {\bf 39S1}, 9 (2005),
  {\tt hep-lat/0211003}.

\bibitem{Jpsi-lattice2}
  M.~Asakawa and T.~Hatsuda,
  {\it $J/\psi$ and $\eta_c$ in the deconfined plasma from lattice QCD,}
  Phys.\ Rev.\ Lett.\  {\bf 92}, 012001 (2004),
  {\tt hep-lat/0308034}.
\bibitem{Jpsi-lattice3}
  S.~Datta, F.~Karsch, P.~Petreczky and I.~Wetzorke,
  {\it Behavior of charmonium systems after deconfinement,}
  Phys.\ Rev.\ D {\bf 69}, 094507 (2004),
  {\tt hep-lat/0312037}.

\bibitem{Jpsi-lattice4}
  A.~Jakovac, P.~Petreczky, K.~Petrov and A.~Velytsky,
  {\it On charmonia survival above deconfinement,}
  {\tt hep-lat/0603005}.
 
\bibitem{Shuryak}
  E.~V.~Shuryak and I.~Zahed,
  {\it Rethinking the properties of the quark gluon plasma at $T\sim T_c$,}
  Phys.\ Rev.\ C {\bf 70}, 021901 (2004),
  {\tt hep-ph/0307267}.

\bibitem{wall1}
  E.~Witten,
  {\it Anti-de Sitter space, thermal phase transition,
  and confinement in  gauge theories,}
  Adv.\ Theor.\ Math.\ Phys.\  {\bf 2}, 505 (1998),
  {\tt hep-th/9803131}.

\bibitem{wall2}
For an introduction, see
  J.~M.~Maldacena,
  {\it TASI-2003 lectures on AdS/CFT,}
  {\tt hep-th/0309246}.

\bibitem{wall3}
For a phenomenological application, see
  J.~Erlich, E.~Katz, D.~T.~Son and M.~A.~Stephanov,
  {\it QCD and a holographic model of hadrons,}
  Phys.\ Rev.\ Lett.\  {\bf 95}, 261602 (2005),
  {\tt hep-ph/0501128}.

\bibitem{chris}
  C.~P.~Herzog,
  {\it A holographic prediction of the deconfinement temperature,}
  {\tt hep-th/0608151}.

\bibitem{AMY2}
P.~Arnold, G.~D.~Moore and L.~G.~Yaffe,
  {\it Photon emission from ultrarelativistic plasmas,}
  JHEP {\bf 0111}, 057 (2001),
  {\tt hep-ph/0109064}.

\bibitem{Baier}
R.~Baier, H.~Nakkagawa, A.~Niegawa and K.~Redlich,
  {\it Production rate of hard thermal photons and screening of quark mass
  singularity,}
  Z.\ Phys.\ C {\bf 53}, 433 (1992).

\bibitem{Kapusta}
J.~I.~Kapusta, P.~Lichard and D.~Seibert,
  {\it High-energy photons from quark-gluon plasma versus hot hadronic gas,}
  Phys.\ Rev.\ D {\bf 44}, 2774 (1991)
  [Erratum-ibid.\ D {\bf 47}, 4171 (1993)].

\bibitem{Aurenche}
P.~Aurenche, F.~Gelis, R.~Kobes and H.~Zaraket,
  {\it Bremsstrahlung and photon production in thermal {QCD},}
  Phys.\ Rev.\ D {\bf 58}, 085003 (1998),
  {\tt hep-ph/9804224}.

\bibitem{LP}
L.~D.~Landau and I.~Pomeranchuk,
  {\it Limits of applicability of the theory of bremsstrahlung electrons and pair
  production at high-energies,}
  Dokl.\ Akad.\ Nauk Ser.\ Fiz.\  {\bf 92}, 535 (1953).

\bibitem{M1}
A.~B.~Migdal,
  {\it Quantum kinetic equation for multiple scattering,}
  Dokl.\ Akad.\ Nauk Ser.\ Fiz.\  {\bf 105}, 77 (1955).

\bibitem{M2}
  A.~B.~Migdal,
  {\it Bremsstrahlung and pair production in condensed media at high-energies,}
  Phys.\ Rev.\  {\bf 103}, 1811 (1956).

\bibitem{Braaten}
E.~Braaten and R.~D.~Pisarski,
  {\it Soft amplitudes in hot gauge theories: A general analysis,}
  Nucl.\ Phys.\ B {\bf 337}, 569 (1990).

\bibitem{Gelis}
P.~Aurenche, F.~Gelis and H.~Zaraket,
  {\it A simple sum rule for the thermal gluon spectral function and
  applications,}
  JHEP {\bf 0205}, 043 (2002),
  {\tt hep-ph/0204146}.

\bibitem{MooreRobert}
G.~D.~Moore and J.~M.~Robert,
  {\it Dileptons, spectral weights, and conductivity in the quark-gluon
    plasma,}
  {\tt hep-ph/0607172}.

\bibitem{AMY1}
P.~Arnold, G.~D.~Moore and L.~G.~Yaffe,
  {\it Transport coefficients in high temperature gauge theories. I:
  Leading-log results,}
  JHEP {\bf 0011}, 001 (2000),
  {\tt hep-ph/0010177}.

\bibitem{AMY4}
P.~Arnold, G.~D.~Moore and L.~G.~Yaffe,
{\it Transport coefficients in high temperature gauge theories. II: Beyond
leading log,}
JHEP {\bf 0305}, 051 (2003),
{\tt hep-ph/0302165}.

\bibitem{Jan-e}
  Jan-e Alam, private communication.

\bibitem{Majumder}
A.~Majumder and C.~Gale,
  {\it On the imaginary parts and infrared divergences of two-loop
    vector  boson self-energies in thermal QCD,}
  Phys.\ Rev.\ C {\bf 65}, 055203 (2002),
  {\tt hep-ph/0111181}.

\bibitem{N=2*}
  A.~Buchel and J.~T.~Liu,
  {\it Thermodynamics of the $\mathcal N\,{=}\,2$* flow,}
  JHEP {\bf 0311}, 031 (2003),
  {\tt hep-th/ 0305064}.

\bibitem{Buchel:2004di}
  A.~Buchel, J.~T.~Liu and A.~O.~Starinets,
  {\it Coupling constant dependence of the shear viscosity in $\Nfour$
  supersymmetric Yang-Mills theory,}
  Nucl.\ Phys.\ B {\bf 707}, 56 (2005)
  {\tt hep-th/0406264}.

\bibitem{KarchKatz}
  A.~Karch and E.~Katz,
  {\it Adding flavor to AdS/CFT,}
  JHEP {\bf 0206}, 043 (2002)
  {\tt hep-th/0205236}.





\bibitem{hep-th/0104065}
  G.~Policastro and A.~Starinets,
  {\it On the absorption by near-extremal black branes,}
  Nucl.\ Phys.\ B {\bf 610}, 117 (2001),
  {\tt hep-th/0104065}.

\bibitem{olver}
  F.~W.~J.~Olver, {\it Asymptotics and special functions},
  A\, K\, Peters, Wellesley, 1997.

\end{thebibliography}
\end{document}